\documentclass[subscriptcorrection,upint,varvw,barcolor=Goldenrod3,mathalfa=cal=euler,balance,hyphenate,pdf-a,nofoot,nolists]{asmejour} %

\usepackage{graphicx} 
\usepackage{hyperref}   
\hypersetup{
	colorlinks=true,
	linkcolor=blue,
	citecolor=blue,
	urlcolor=blue,
}

\usepackage{siunitx}
\usepackage{subcaption}

\newcommand{\etal}{\textit{et al.}}
\usepackage{stfloats}
\usepackage{enumitem}
\usepackage{tabularx}
\usepackage{array}
\usepackage{booktabs}
\usepackage{mdframed}
\usepackage{graphicx}
\usepackage{graphicx}
\usepackage{colortbl}
\usepackage{enumitem}

\newcolumntype{L}{>{\small\bfseries\raggedright\arraybackslash}p{3cm}}

\newmdenv[
  topline=false,
  bottomline=false,
  rightline=false,
  linewidth=2pt,
  linecolor=blue,
  backgroundcolor=gray!20,
  leftmargin=10pt,
  rightmargin=10pt,
  innertopmargin=10pt,
  innerbottommargin=10pt
]{customquote}

\begin{document}

\JourName{}

\SetAuthorBlock{Mohammadmehdi Ataei\CorrespondingAuthor}{%
Autodesk Research, \\
661 University Avenue,\\
Toronto, Ontario M5G 1M1, Canada\\
email: mehdi.ataei@autodesk.com
}

\SetAuthorBlock{Hyunmin Cheong}{%
Autodesk Research, \\
661 University Avenue,\\
Toronto, Ontario M5G 1M1, Canada
}

\SetAuthorBlock{Daniele Grandi}{%
Autodesk Research, \\
The Landmark @ One Market, Ste. 400,\\
San Francisco, CA 94105, USA
}

\SetAuthorBlock{Ye Wang}{%
Autodesk Research, \\
The Landmark @ One Market, Ste. 400,\\
San Francisco, CA 94105, USA
}

\SetAuthorBlock{Nigel Morris}{%
Autodesk Research, \\
661 University Avenue,\\
Toronto, Ontario M5G 1M1, Canada
}

\SetAuthorBlock{Alexander Tessier}{%
Autodesk Research, \\
661 University Avenue,\\
Toronto, Ontario M5G 1M1, Canada
}

\title{Elicitron: An LLM Agent-Based Simulation Framework for Design Requirements Elicitation}

\keywords{Requirement Elicitation, Large Language Models, Artificial Intelligence, LLM Agents, Computer Aided Design}

\begin{abstract}

Requirements elicitation, a critical, yet time-consuming and challenging step in product development, often fails to capture the full spectrum of user needs. This may lead to products that fall short of expectations. This paper introduces a novel framework that leverages Large Language Models (LLMs) to automate and enhance the requirements elicitation process.\  LLMs are used to generate a vast array of simulated users (LLM agents), enabling the exploration of a much broader range of user needs and unforeseen use cases.\ These agents engage in product experience scenarios, through explaining their actions, observations, and challenges.\ Subsequent agent interviews and analysis uncover valuable user needs, including latent ones. We validate our framework with three experiments.\ First, we explore different methodologies for diverse agent generation, discussing their advantages and shortcomings.\ We measure the diversity of identified user needs and demonstrate that context-aware agent generation leads to greater diversity.\ Second, we show how our framework effectively mimics empathic lead user interviews, identifying a greater number of latent needs than conventional human interviews.\ Third, we showcase that LLMs can be used to analyze interviews, capture needs, and classify them as latent or not. Our work highlights the potential of using LLM agents to accelerate early-stage product development, reduce costs, and increase innovation. 

\end{abstract}

\date{}
\maketitle







\section{Introduction}

Requirements elicitation (RE) sits at the core of successful product design, yet it remains a complex and resource-intensive endeavor.\ Traditional RE methods, like interviews, focus groups, and prototyping, are invaluable but have inherent limitations. These methods are often time-consuming, may not fully capture the diversity of user perspectives, and may miss underlying needs that are difficult for users to articulate~\cite{zave1997classification,berry2007ambiguity}.\ The consequences of inadequate requirements elicitation can be significant, ranging from design misalignment to compromised product adoption.

Recent advancements in Large Language Models (LLMs) present new possibilities for automating and augmenting requirements elicitation.\ LLMs, having learned the patterns and complexities of human language from vast textual corpora, seemingly possess a remarkable capacity for natural language understanding~\cite{brown2020language}. This potential can be leveraged to construct a simulated environment where LLM agents \emph{role-play} a variety of potential users.\ These agents can embody distinct viewpoints, engage in product experience scenarios, and participate in user interviews aimed at identifying user needs.

This research presents a new LLM-based framework, called \textit{Elicitron}, for automating and augmenting the RE process.\ In Elicitron, LLM agents are constrained to produce structured outputs that are relevant to and useful for the requirement elicitation process.\ Elicitron also employs techniques to create diverse user-representing agents and simulate product experiences through the \emph{Action, Observation, Challenge} steps, inspired by \emph{chain-of-thought} reasoning~\cite{wei2022chain}.\ Lastly, by creating agents with specific roles, Elicitron can discover interesting user needs that may otherwise be difficult to obtain with human interviews.

Elictron's ability to create diverse user agents can be leveraged to identify a diverse set of user needs.\ Because the process of creating and interviewing these agents is automated, the process is highly scalable, unlike the traditional RE methods.\ We conducted an experiment to evaluate the capability of Elicitron to generate a diverse set of user needs and identified a context-aware generation method that maximizes needs diversity. 

In addition, we show how Elicitron can be applied to identify latent needs -- those unarticulated and unexpected factors that strongly influence product desirability~\cite{maalej2015toward} -- which is significantly difficult to obtain with the traditional RE methods.\ This can be achieved by either automatically or manually creating user agents with \emph{empathic lead user} roles~\cite{lin2007}.\ We conduct a second experiment to demonstrate that Elicitron can generate a higher number of latent needs than human interviews. Our third experiment shows that, given a criteria and chain-of-thought reasoning, LLMs are capable to identifying and classifying latent needs within interview data.


\subsection{LLMs for Diverse and Latent Needs Identification}

There are reasons to believe LLMs could offer unique advantages for identifying diverse and latent needs.\ The core capabilities of LLMs lie in their flexibility and ability to perform different tasks related to natural language.\ Their inclusion of contextual elements greatly improves their performance and result in them appearing to pick up on nuances and inferences.\ Because LLMs have been trained on a vast amount of data, they likely have been exposed to a diverse set of user needs for a particular product.\ During training, LLM's may have also picked up behavioral patterns and subtle language cues that hint at underlying--sometimes even subconscious--needs experienced by users.\ Moreover, it may be able to perform analogical reasoning to relate experiences and needs identified across different products and uncover novel needs. 


However, a potential contradiction arises when considering the inherent nature of LLMs: They are trained to predict the most likely next word or sequence of words based on patterns in their training data~\cite{vaswani2017attention}. On the surface, this focus on selecting the most likely outcome might seem at odds with the goal of uncovering diverse or latent needs.\ In practice, LLMs do not always select the next word or phrase greedily due to certain hyper-parameters (such as temperature or top-P) allowing for some randomness.\ Furthermore, the context provided to the LLM, in the form of input text that the model considers when generating a response, can heavily influence its output.\ Through careful contextualization, we can encourage the LLM to move beyond the obvious outputs and for our purpose, produce outputs that help the designer discover diverse and latent needs.

\section{Background}

\begin{figure*}[ht]
    \centering
    \includegraphics[width=1.0\textwidth]{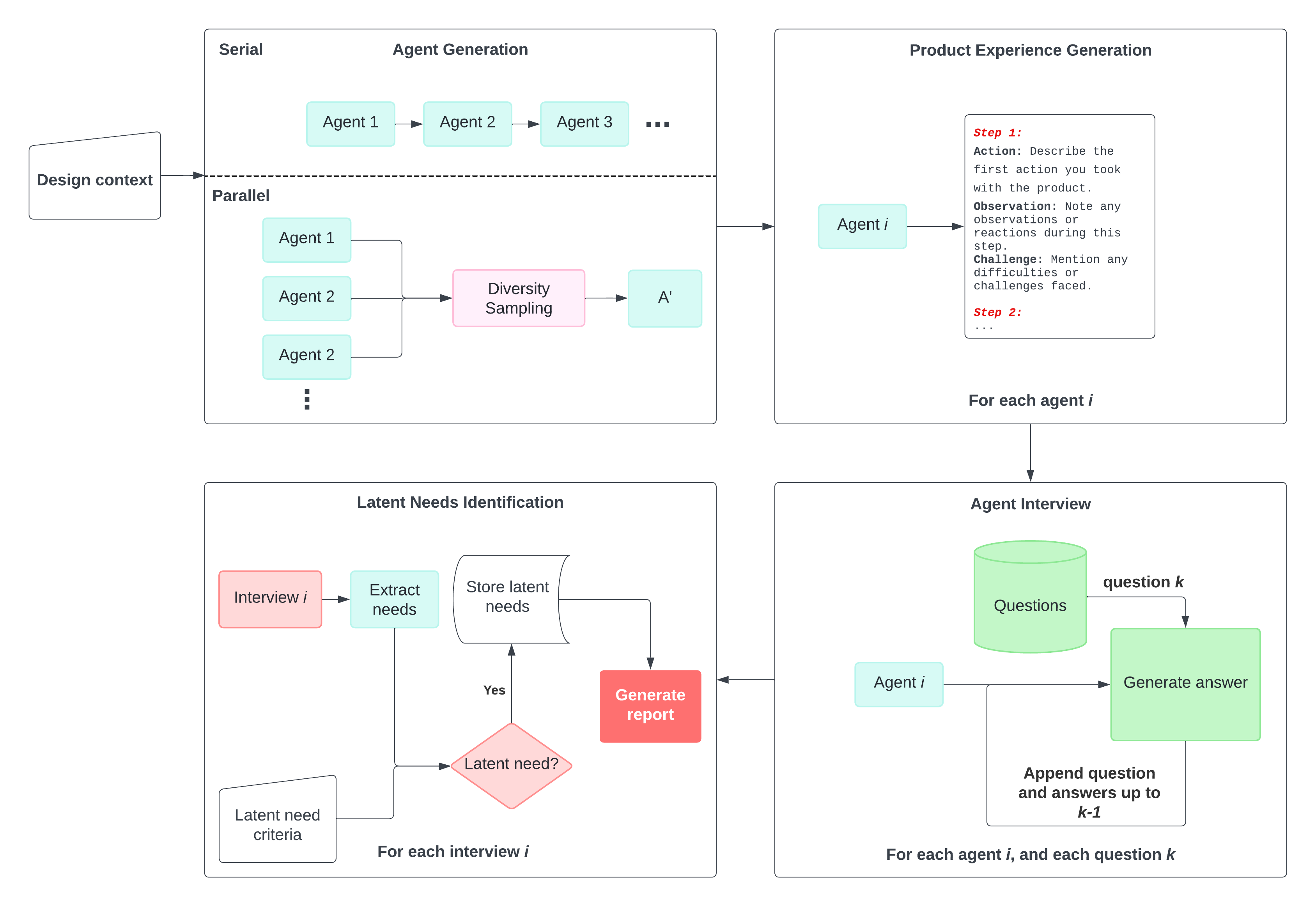}
    \caption{Elictron's architecture for requirements elicitation using LLMs: First, LLM agents are generated within a design context in either serial and parallel fashion (incorporating diversity sampling to represent varied user perspectives). These agents then engage in simulated product experience scenarios, documenting each step (Action, Observation, Challenge) in detail.\ Following this, they undergo an agent interview process, where questions are asked and answered to surface latent user needs.\ In the final stage, latent needs are identified using an LLM on a provided criteria, and finally a report is generated from the identified latent needs.
}
    \label{fig:architecture}
\end{figure*}

\subsection{Large Language Models}

LLMs are machine learning models that appear to exhibit the ability to understand, reason with, and generate natural language~\cite{brown2020language}.\ LLMs have been shown to engage in fluent conversations, translate languages, and write various types of creative content in different styles.\ The development of LLMs marks a significant step towards human-like agent capabilities within artificial intelligence.\ Beyond traditional applications, the use of LLMs is expanding into areas like software development, content creation, and customer service~\cite{lingard2023writing,htet2024chatgpt,subagja2023improving,li2022competition}. 

LLMs are trained using self-supervised learning.\ They are trained on vast amounts of text data and to predict the next word or sequence of words in a sentence. This training process, along with architectural advancements like the Transformer model~\cite{vaswani2017attention}, allow LLMs to grasp patterns, nuances, and context within human language, including learning implied meanings~\cite{brown2020language}.

The knowledge acquired during training provides the foundation for LLMs to effectively engage in role-playing scenarios~\cite{shanahan2023role,csepregi2021effect,zhu2023calypso}.\ They can simulate various roles by using specific language patterns, word choices, and sentence structures learned from their training data.\ For example, if an LLM encounters a training dataset rich with technical communication, it can adapt its vocabulary and sentence structure to convincingly adopt the role of an engineer during role-playing sessions.

LLMs' capability to role-play allows them to become useful tools for use in requirements elicitation.\ They can simulate a diverse set of users, including empathic lead users, providing valuable insights into their product experience and needs.

\subsection{Design Requirement Elicitation, Empathic Design, and Latent Needs}
During the requirement elicitation phase of design engineering, empathy plays an important role in helping engineers better understand user needs~\cite{gray2015,schmitt2016}, and develop a better understanding of the design problem~\cite{walther2012}.\ By interviewing, observing, and empathizing with users, designers can derive structured design requirements from unstructured feedback from product users, in a process called \emph{empathic design}~\cite{hannukainen2006,Leonard1997}. 

There are two types of design requirements or user needs: direct needs, which tend to be obvious to the customer and lead to incremental changes in a product, and latent needs, which may be non-obvious and difficult to uncover~\cite{Otto2003}.

User interviews or observations may not reveal latent needs that consumers deem important in a final product~\cite{Ulrich2016}.\ Identifying latent needs early in the design process has been found to speed up the development process, and their discovery benefits the design engineer by providing insights into extreme use cases that might push the product to its limits~\cite{Suh1990,vonhippel1986,urban1988}. 

Over the years, the design research community has experimented with various empathic design methods to improve latent need discovery.\ Hannukainen and Holtta-Otto used photo dairy and contextual inquiry with disabled people to identify latent needs~\cite{hannukainen2006}.\ Lin~\etal~elicited latent needs from ordinary users by simulating extraordinary situations, e.g., using a blindfold to simulate limited sight or oven mittens to simulate limited dexterity~\cite{lin2007}.\ Issa~\etal~prompted designers interpreting user interviews to ``write a statement as someone with [specific experience]'', in an attempt to bias the designer to be more empathetic with a lead user~\cite{issa2023}.\ More recently, Zhu~\etal~theorized how artificial intelligence and LLMs might be leveraged to support data-driven user studies for empathic design~\cite{zhu2023}. 

While the design research community has identified empathy as an important component of design requirement elicitation~\cite{lin2007,strobel2013,raviselvam2017,surma-aho2018,tang2018}, interviewing empathic lead users or observing them in user studies remains a time-consuming and costly activity.\ The use of LLMs to simulate this process by creating and interviewing a diverse set of empathic lead user agents could address the gaps.

\subsection{Metrics for Design Diversity}
\label{sec:diversity}
In design methodology literature, design diversity is typically considered the extension of novelty from a set of designs.\ A novel design is often considered to be unique by the person who created it (psychological novelty), or more generally unique to the field (historical novelty)~\cite{boden2009computer}.\ Evaluating the novelty of a design is a subjective task, and it is typical to leverage Consensual Assessment Technique (CAT) or similar methods that involve asking domain experts to rate designs on criteria such as novelty~\cite{amabile1988model, amabile1982social, miller2021should, amabile2018creativity}.\ However, recent work in deep generative models, which leverages machine learning methods trained on large datasets of prior designs to create novel design solutions, has led to the development and adoption of computational novelty and diversity metrics.\ These include the convex hull volume and the mean distance to centroid metrics, which can be used to measure the average diversity of a whole set of designs~\cite{regenwetter2023statistical, ma2023conceptual,jiralerspong2024feature,sarica2023innovation,picard2023dated,regenwetter2023counterfactuals,bagazinski2023shipgen,fan2023noise}.\ In addition to those measures, we are also interested in measuring the diversity of possible clusters of design ideas, and not just the outliers, thus the \textit{silhouette score}  typically used in cluster analysis~\cite{rousseeuw1987silhouettes} may also be applicable.

The \textit{convex hull volume} is defined as the hypervolume of the smallest convex set that includes all of the samples.\ It has been used to measure the diversity in different disciplines, but is sensitive to outliers~\cite{regenwetter2023statistical, ma2023conceptual, podani2009convex}.\ A larger convex hull volume indicates that, in the embedding space, the samples cover more space and are more diverse. 

The \textit{mean distance to centroid} of the embeddings is computed by taking the mean of each sample to the centroid of the whole set~\cite{regenwetter2023statistical, ma2023conceptual, mueller2015combining, brown2019quantifying}.\ A larger value indicates that the samples in the set are further from the centroid, and thus are assumed to represent more diverse concepts.\ This metric is more suitable for more uniform distributions, and is also sensitive to outliers. 

In the context of clustering algorithms, the \textit{silhouette score} is a metric used to calculate the performance of a clustering technique.\ Its value ranges from -1 to 1, where a high value indicates that the object is well-matched to its own cluster and poorly matched to neighboring clusters.\ The silhouette score has been used to measure the diversity of recommendation systems, but has not been adopted in the field of design research~\cite{zanitti2022exploiting, chaudhuri2019advanced}.

\section{Architecture of Elicitron}
\label{sec:architecture}
Elicitron is designed to closely simulate real-world requirements elicitation processes.\ The architecture comprises four distinct components mirroring the phases involved in gathering requirements (Figure \ref{fig:architecture}).\ Each component is discussed in detail below.\ 

To maintain structural integrity and prevent workflow errors, we employ a Pydantic model to shape LLM outputs, followed by a schema validation step.

\subsection{Agent Generation}

A significant challenge in requirements elicitation (either with LLMs or traditional methods) is capturing a diverse range of user viewpoints.\ For this reason, our framework's initial step generates a diverse set of agents to simulate users within the elicitation process.\ This mirrors real-world practices where a wide variety of users are deliberately selected for RE studies.

The LLM is instructed to generate three elements for each user agent:
\begin{itemize}[noitemsep]
    \item \textbf{Name:} A label representing the user agent.
    \item \textbf{Description:} A description of the user characteristics.
    \item \textbf{Reasoning Chain:} A rationale for creating this agent.
\end{itemize}
The first two elements comprise the description of a user role.\ The reasoning chain aids in understanding the LLM's agent generation logic, a process similar to \emph{chain-of-thought}~\cite{wei2022chain}.

We have employed the following agent generation methods:

\subsubsection{Parallel Agent Generation}

Here, the LLM receives N independent prompts to generate N user agents simultaneously. This method is advantageous for rapid creation of a large number of agents in parallel. However, due to the LLM's lack of awareness of other agents being generated, diversity may be limited as the model could produce similar agents.

To mitigate this, we implement a \emph{filtering} stage.\ We use a KMeans clustering algorithm to group agents based on the similarity of their embeddings and then select only representative agents from each cluster to result in a diverse set of agents. 

Given a set of generated agents $A = {a_1, a_2, \ldots, a_N}$, where each agent $a_i$ is represented by an embedding vector $v_i$ in a high-dimensional space, the goal is to select a diverse subset of agents.

\begin{enumerate}[noitemsep]
\item \textbf{Assign Embeddings:} First, assign an embedding vector $v_i$ to each agent $a_i$ description in the set $A$.\ We used \texttt{text-embedding-ada-002} by OpenAI for this purpose.

\item \textbf{Perform KMeans Clustering:} Apply a clustering algorithm, KMeans$(V, k)$.\ $V$ is the matrix of all embedding vectors $v_i$ and $k$ is a chosen number of clusters (which is less than or equal to $N$).\ Given $k$, the KMeans algorithm assigns each agent to the cluster with the nearest mean embedding.

\item \textbf{Select Diverse Agents:} From each of the $k$ clusters, select one representative agent.\ The selected agents are deemed diverse, as they come from different clusters in the embedding space.\ The resulting set of agents is denoted by $A'$.

\end{enumerate}

This method may involve overgenerating agents, followed by filtering down to N agents. While we used KMeans for filtering, other clustering techniques are also applicable.

\subsubsection{Serial Agent Generation}

In this technique, the LLM receives a single prompt to generate N agents.\ Here, the details of generated agents persist in the LLM's context to promote greater diversity compared to parallel generation.\ A downside is decreased speed compared to parallel generation, and a theoretical limit on the number of agents generated based on the LLM's maximum token output length.\ At the time of this writing, most LLMs cap output at around 4096 tokens, which experimentally suggests a maximum of roughly 20 agents per call for each generation.

\subsection{Product Experience Generation}

After generating a diverse agent pool, user agents are prompted to hallucinate their interaction with the potential product.\ This phase is essential for identifying specific usage scenarios that could lead to detailed and latent needs to be identified during the subsequent interview process.

\begin{enumerate}
    \item \textbf{Simulated Interaction:} Agents receive an open-ended prompt to describe steps they would take to interact with the product.\ This might involve setup, specific feature usage, or troubleshooting.\ Agents are allowed to explore freely, simulating the varied ways real users would interact with the product.
    
    \item \textbf{Structured Response Generation:} For each interaction step, agents provide responses organized into three elements:

    \begin{itemize}[noitemsep]
        \item \textbf{Action:} The description of the interaction step taken (e.g., setup, feature activation).
        \item \textbf{Observation:} The agent's reactions and perceptions of the step. This includes both favorable impressions and points of friction.
        \item \textbf{Challenge:} Explicit articulation of obstacles or difficulties encountered. This is done to uncover pain points in the user experience.
    \end{itemize}

\end{enumerate}

These structured responses are then utilized as context in the follow-up interview phase. This helps to contextualize the agent's experience, mirroring how chain-of-thought prompting~\cite{wei2022chain} aids in deeper response generation.\ An example of the output of product experience generation can be found in Section~\ref{sec:example_output}.

\subsection{Agent Interview}

The agent interview step mirrors real-world user interviews.\ It prompts each agent to reflect back on their product experience and asks follow-up questions aimed at uncovering user needs and nuanced insights identified from their product experience. 
The process works as follows:

\begin{enumerate}
\item \textbf{Question Pool Creation:} A set of interview questions is prepared (human-developed or automatically generated by an LLM).\ These questions could be tailored to cover multiple aspects of the product, while also asking for innovative insights or improvement ideas.

\item \textbf{Contextualized Questioning:} Questions are asked to each agent, integrating their prior Q\&A responses and simulated product experiences into the LLM's context.\ This contextualization aim to alleviate the LLM's tendency to provide generic responses and facilitates the answers to be based on the individual agent's unique experience.
\end{enumerate}


\subsection{Latent Needs Identification}
In this workflow stage, we leverage previously collected interview responses to isolate needs automatically.\ An LLM processes agent interviews, extracting expressed needs.\ It then provides step-by-step reasoning for each identified need, drawing from established latent need criteria and examples, provided by the human experts.\ Finally, the LLM compiles all findings into a detailed report, offering insights on both expressed and latent needs uncovered during the analysis.

After the responses are collected from all user agents, the designer can review them to identify user needs that could be utilized for the subsequent design process. 

In this work, we conducted two experiments to examine the value of Elicitron, all using \textit{GPT-4-Turbo} from OpenAI as the LLM \cite{openai2024gpt4}.

\section{Experiment 1: Automatic Generation of Diverse Users and Their Needs}

To examine the value of Elicitron in terms of identifying diverse user needs, we evaluated the agent generation methods proposed in Section~\ref{sec:architecture}.\ We generated 20 user agents each using three conditions: serial, parallel, and parallel with the KMeans filtering, and compared the diversity of generated agents and their responses using computational metrics.

\begin{figure*}[t]
    \centering
    \includegraphics[width=\linewidth]{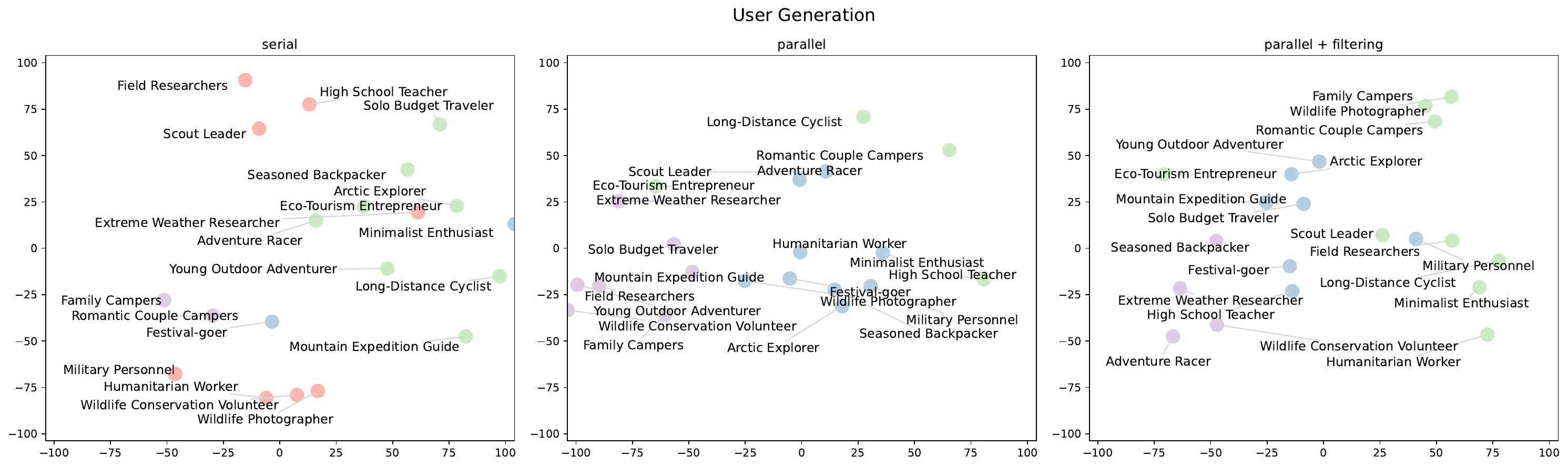}
    \caption{Four groups of users' embeddings after reducing dimensions to 2 using t-SNE.\ Group 1: Service and Conservation \colorbox[RGB]{251,185,185}{(in red)}.\ Group 2: Outdoor Recreation and Camping \colorbox[RGB]{188,203,229}{(in blue)}.\ Group 
    3: Adventure and Exploration \colorbox[RGB]{185, 251, 185}{(in green)}.\ Group 4: Family Camping and Outdoor Activities \colorbox[RGB]{220,200,231}{(in purple)}.\ The serial generation gives the best coverage of all four groups.\ Parallel generation with and without filtering both missed service and conservation-related users.}
    \label{fig:stakeholder_diversity}
\end{figure*}

\subsection{Evaluation of Diversity}
\subsubsection{Computational Evaluation}
As discussed in Section \ref{sec:diversity}, there is no de facto method to computationally evaluate the diversity of interview participants and their corresponding responses.\ Thus, we leveraged three different methods used in prior work that rely on embedding the generated responses in a latent space to measure the diversity of design solutions. 

For all three metrics, we first generated embeddings for each of the role descriptions and responses to 12 interview questions using \textit{text-embedding-ada-002}.\ Note the interview questions asked are presented in Section 5.1. This resulted in 13 sets of 20 embeddings (for the user role and for each of the 12 questions). For each of our three methods used to generate the data (`serial', `parallel', and `parallel with filtering') we then computed the \textit{convex hull volume} and the \textit{mean distance to centroid} from the individual sets of 20 embeddings, and normalize the results from 0 to 1.

The \textit{silhouette score} is defined as follows.\ If $a$ is the mean intra-cluster distance (the average distance between each point within a cluster), and $b$ is the mean nearest-cluster distance (the average distance from a point to the nearest cluster of which it is not a part), then the silhouette score $s$ for a single sample is given by the formula $s = \frac{b - a}{\max(a, b)}$.
The mean silhouette score for a set of samples is the mean of the individual silhouette scores for each sample.\ An appropriate number of $k$ clusters can then be selected by choosing the $k$ cluster with the highest mean silhouette score, indicating that each cluster is very compact and distinct from other clusters.\ We used this metric both to choose a number of clusters for KMeans, as well as a standalone metric to quantify the diversity of the samples for each of our methods.

\subsubsection{Qualitative Evaluation}
\label{qualitative}
To further understand the differences along the diversity, we evaluated the content of role descriptions and interview responses as follows:
\begin{enumerate}[noitemsep]
    \item Cluster the embeddings of all 60 user agents generated from the three conditions using KMeans.\ $k$ is chosen based on the silhouette score to maximize distinct clusters.  
    \item Summarize the $k$ clusters of agents with the LLM with the following prompt, ``Here are $k$ groups of users, give a theme for each group.\ Group 1: ...''. For the content of each group, the role descriptions of the agents in each cluster are used.
    \item Examine the coverage of clusters for each condition using scatter plots after reducing the dimensions with t-distributed stochastic neighbor embedding (t-SNE), shown in Figure \ref{fig:stakeholder_diversity}~\cite{van2008visualizing}.
\end{enumerate}

\subsection{Results}

\subsubsection{Convex Hull}
Table~\ref{tab:convex} shows that the convex hull volumes were higher for the serial method than the parallel and parallel with filter methods, indicating a significant increase in diversity. Also, on average, the filtering method improved the diversity of the parallel generation method.

\begin{table}[h]
\centering
\resizebox{0.9\columnwidth}{!}{%
\begin{tabular}{@{}rlll@{}}
\toprule
\textbf{} & \multicolumn{1}{c}{\textbf{Serial}} & \multicolumn{1}{c}{\textbf{Parallel}} & \multicolumn{1}{c}{\textbf{\begin{tabular}[c]{@{}c@{}}Parallel \\ + filtering\end{tabular}}} \\ \midrule
\textbf{User} & \cellcolor[HTML]{023858}{\color[HTML]{F1F1F1} 0.991878} & \cellcolor[HTML]{FDF5FA}0.097886 & \cellcolor[HTML]{FFF7FB}0.081218 \\
\textbf{Characteristics} & \cellcolor[HTML]{023858}{\color[HTML]{F1F1F1} 0.928448} & \cellcolor[HTML]{FFF7FB}0.191309 & \cellcolor[HTML]{E1DFED}0.318411 \\
\textbf{Size} & \cellcolor[HTML]{023858}{\color[HTML]{F1F1F1} 0.717999} & \cellcolor[HTML]{FFF7FB}0.382065 & \cellcolor[HTML]{4496C3}{\color[HTML]{F1F1F1} 0.581811} \\
\textbf{Shape} & \cellcolor[HTML]{023858}{\color[HTML]{F1F1F1} 0.929433} & \cellcolor[HTML]{FFF7FB}0.247194 & \cellcolor[HTML]{F9F2F8}0.273950 \\
\textbf{Weight} & \cellcolor[HTML]{023858}{\color[HTML]{F1F1F1} 0.789456} & \cellcolor[HTML]{FFF7FB}0.314839 & \cellcolor[HTML]{89B1D4}0.526912 \\
\textbf{Material} & \cellcolor[HTML]{023858}{\color[HTML]{F1F1F1} 0.769544} & \cellcolor[HTML]{FFF7FB}0.310822 & \cellcolor[HTML]{62A2CB}{\color[HTML]{F1F1F1} 0.557846} \\
\textbf{Safety} & \cellcolor[HTML]{023858}{\color[HTML]{F1F1F1} 0.659648} & \cellcolor[HTML]{FFF7FB}0.405045 & \cellcolor[HTML]{045483}{\color[HTML]{F1F1F1} 0.633090} \\
\textbf{Durability} & \cellcolor[HTML]{023858}{\color[HTML]{F1F1F1} 0.910748} & \cellcolor[HTML]{E0DDED}0.347797 & \cellcolor[HTML]{FFF7FB}0.222656 \\
\textbf{Aesthetics} & \cellcolor[HTML]{023858}{\color[HTML]{F1F1F1} 0.944532} & \cellcolor[HTML]{FFF7FB}0.152274 & \cellcolor[HTML]{E1DFED}0.290983 \\
\textbf{Ergonomics} & \cellcolor[HTML]{023858}{\color[HTML]{F1F1F1} 0.861179} & \cellcolor[HTML]{FFF7FB}0.338084 & \cellcolor[HTML]{F3EDF5}0.379566 \\
\textbf{Cost} & \cellcolor[HTML]{023858}{\color[HTML]{F1F1F1} 0.910723} & \cellcolor[HTML]{FFF7FB}0.172481 & \cellcolor[HTML]{C8CDE4}0.375278 \\
\textbf{Setup} & \cellcolor[HTML]{023858}{\color[HTML]{F1F1F1} 0.925232} & \cellcolor[HTML]{F4EEF6}0.292025 & \cellcolor[HTML]{FFF7FB}0.242214 \\
\textbf{Transport} & \cellcolor[HTML]{023858}{\color[HTML]{F1F1F1} 0.944723} & \cellcolor[HTML]{FFF7FB}0.222418 & \cellcolor[HTML]{FBF4F9}0.240892 \\ \midrule
\textbf{Mean} & \cellcolor[HTML]{023858}{\color[HTML]{F1F1F1} 0.867965} & \cellcolor[HTML]{FFF7FB}0.267249 & \cellcolor[HTML]{E5E1EF}0.363448 \\ \bottomrule
\end{tabular}%
}
\caption{Convex hull volumes of the embeddings of user role/descriptions and responses to interview questions for each generation method, normalized from 0 to 1. A higher convex hull volume indicates a relatively more diverse set.}
\label{tab:convex}
\end{table}


\subsubsection{Mean Distance to Centroid}
Table \ref{tab:centroid} shows the mean distance to centroid values.\ The results show that the serial method led to more diversity than the parallel and parallel with filtering methods, but by a smaller margin than the convex hull metric.\ Again, the `filtering' method improved the diversity of the `parallel' generation by a small amount. 

\begin{table}[h]
\centering
\resizebox{0.9\columnwidth}{!}{%
\begin{tabular}{@{}rlll@{}}
\toprule
\multicolumn{1}{c}{\textbf{}} & \multicolumn{1}{c}{\textbf{Serial}} & \multicolumn{1}{c}{\textbf{Parallel}} & \multicolumn{1}{c}{\textbf{\begin{tabular}[c]{@{}c@{}}Parallel \\ + filtering\end{tabular}}} \\ \midrule
\textbf{User} & \cellcolor[HTML]{023858}{\color[HTML]{F1F1F1} 0.660156} & \cellcolor[HTML]{FFF7FB}0.527555 & \cellcolor[HTML]{F7F0F7}0.534677 \\
\textbf{Characteristics} & \cellcolor[HTML]{023858}{\color[HTML]{F1F1F1} 0.618368} & \cellcolor[HTML]{FFF7FB}0.542512 & \cellcolor[HTML]{B0C2DE}0.568596 \\
\textbf{Size} & \cellcolor[HTML]{023858}{\color[HTML]{F1F1F1} 0.590934} & \cellcolor[HTML]{FFF7FB}0.552930 & \cellcolor[HTML]{03517E}{\color[HTML]{F1F1F1} 0.587423} \\
\textbf{Shape} & \cellcolor[HTML]{023858}{\color[HTML]{F1F1F1} 0.618861} & \cellcolor[HTML]{FFF7FB}0.551452 & \cellcolor[HTML]{EDE8F3}0.559385 \\
\textbf{Weight} & \cellcolor[HTML]{023858}{\color[HTML]{F1F1F1} 0.610335} & \cellcolor[HTML]{FFF7FB}0.543569 & \cellcolor[HTML]{78ABD0}{\color[HTML]{F1F1F1} 0.576215} \\
\textbf{Material} & \cellcolor[HTML]{023858}{\color[HTML]{F1F1F1} 0.601980} & \cellcolor[HTML]{FFF7FB}0.546090 & \cellcolor[HTML]{2A88BC}{\color[HTML]{F1F1F1} 0.582585} \\
\textbf{Safety} & \cellcolor[HTML]{023858}{\color[HTML]{F1F1F1} 0.584500} & \cellcolor[HTML]{FFF7FB}0.562831 & \cellcolor[HTML]{023858}{\color[HTML]{F1F1F1} 0.584449} \\
\textbf{Durability} & \cellcolor[HTML]{023858}{\color[HTML]{F1F1F1} 0.623426} & \cellcolor[HTML]{A2BCDA}0.569565 & \cellcolor[HTML]{FFF7FB}0.535664 \\
\textbf{Aesthetics} & \cellcolor[HTML]{023858}{\color[HTML]{F1F1F1} 0.649814} & \cellcolor[HTML]{FFF7FB}0.521597 & \cellcolor[HTML]{D2D2E7}0.552883 \\
\textbf{Ergonomics} & \cellcolor[HTML]{023858}{\color[HTML]{F1F1F1} 0.614401} & \cellcolor[HTML]{EEE9F3}0.561247 & \cellcolor[HTML]{FFF7FB}0.554539 \\
\textbf{Cost} & \cellcolor[HTML]{023858}{\color[HTML]{F1F1F1} 0.622532} & \cellcolor[HTML]{FFF7FB}0.536028 & \cellcolor[HTML]{9EBAD9}0.570199 \\
\textbf{Setup} & \cellcolor[HTML]{023858}{\color[HTML]{F1F1F1} 0.632032} & \cellcolor[HTML]{F0EAF4}0.552557 & \cellcolor[HTML]{FFF7FB}0.543338 \\
\textbf{Transport} & \cellcolor[HTML]{023858}{\color[HTML]{F1F1F1} 0.633343} & \cellcolor[HTML]{FFF7FB}0.543400 & \cellcolor[HTML]{F2ECF5}0.550992 \\ \midrule
\textbf{Mean} & \cellcolor[HTML]{023858}{\color[HTML]{F1F1F1} 0.620052} & \cellcolor[HTML]{FFF7FB}0.547026 & \cellcolor[HTML]{DBDAEB}0.561611 \\ \bottomrule
\end{tabular}%
}
\caption{Mean distances to the centroid of the embeddings of user role/descriptions and responses to interview questions for each generation method, normalized from 0 to 1.\ A higher mean distance indicates a relatively more diverse set.}
\label{tab:centroid}
\end{table}


\subsubsection{Silhouette Score}
\begin{figure}[h]
    \centering
    \includegraphics[width=0.9\linewidth]{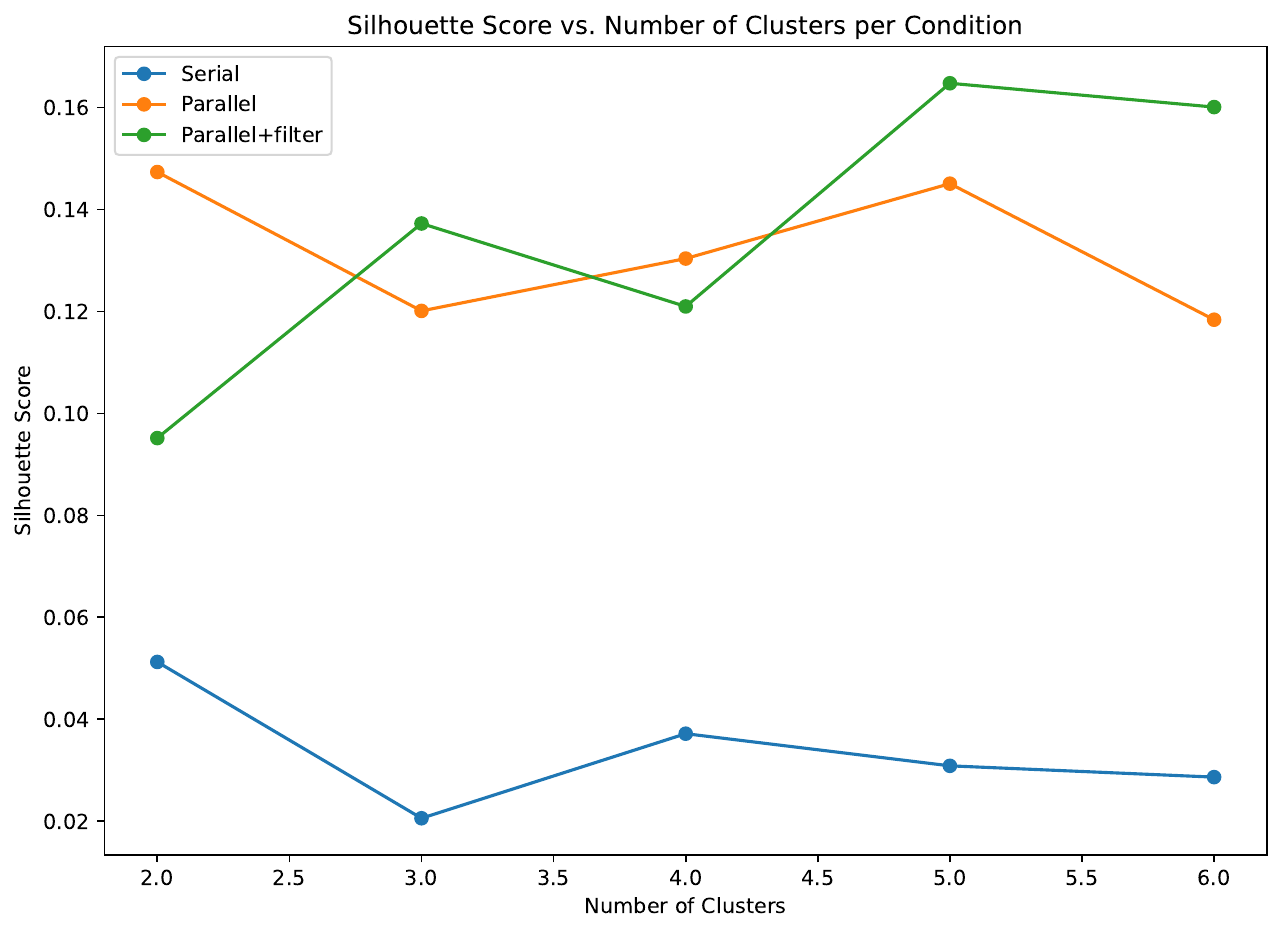}
    \caption{The silhouette score measures the intra- and inter-cluster distance.\ The serial method results in stakeholder embeddings that are more difficult to cluster compared to the parallel and parallel with filtering methods, which indicates that the serial embeddings are more diverse.}
    \label{fig:silhouette}
\end{figure}

We expect that a high silhouette score across various $k$'s would indicate that the points are easier to cluster, or closer to each other, while a low silhouette score would indicate that the points are further apart, less clustered, and thus would represent a more diverse set.\ Figure~\ref{fig:silhouette} shows the computed silhouette scores for the three-generation methods.\ We can infer once again that the serial method produced the most diverse agents, compared to the parallel and parallel with filtering methods.\ There was no clear distinction between the `parallel' methods.

\subsubsection{Qualitative evaluation}

We evaluated the generated users' descriptions to see what categories of users were created.\ We aggregated all 60 users' embeddings created from the three-generation methods. We then chose $k=4$ based on the silhouette score, i.e., when $k=4$ it gives the most distinct clusters.\ Using KMeans with $4$ clusters, we found the following groups of users:  

\begin{enumerate}[noitemsep]
    \item \textbf{Service and Conservation:} Roles including military, humanitarian work, and field research. 
    \item \textbf{Outdoor Recreation and Camping:} Roles with activities such as minimalist camping and stargazing.
    \item \textbf{Adventure and Exploration:} Roles with activities such as hiking, backpacking, and mountain climbing.
    \item \textbf{Family Camping and Outdoor Activities:} Roles emphasizing bonding through camping and nature activities. 
\end{enumerate}

It should be noted that both the parallel and parallel with filtering methods did not create any user roles belonging to the service and conservation groups. 


\subsection{Discussion}
From both computational and qualitative evaluation, the serial generation method led to the most diversity in the user roles and the responses to interview questions.\ The parallel generation method with KMeans filtering helps improve the diversity than using parallel generation only, albeit not significantly.\ The serial generation method benefits from maintaining previous agent generations as additional context to the LLM, allowing the LLM to avoid generating repetitions.  

\section{Experiment 2: Automatic Generation of Latent User Needs}

In the second experiment, Elicitron was employed to automatically generate latent needs, which are difficult to identify using traditional interviews with human users.  

To evaluate our method, the tent design example from~\cite{lin2007} along with their reported results was used.\ In particular, we set our base condition using the \emph{empathic lead user} (ELU) interview technique. This involves simulating extraordinary lead user conditions with regular users and interviewing them as the baseline condition. We then compare the number of latent needs identified with the ELU interview technique versus using three different conditions of our requirements elicitation method.

\begin{table*}[t]
  \fontsize{9pt}{8pt}
  \renewcommand{\arraystretch}{1.0}
  \centering
  \begin{tabular}{lll}
  \toprule
    \textbf{Condition 1 (Automatic)} & \textbf{Condition 2 (Automatic with Steering)} & \textbf{Condition 3 (Manual ELUs)} \\
    \midrule
    Young Outdoor Adventurer & Adventure-seeking Teen & Outdoor enthusiast in the mountains \\
    Family Campers & Retired Nature Enthusiast & Hunter \\
    Seasoned Backpacker & Person with Physical Disability & Camper at desert canyons \\
    Festival-goer & Winter Camper & Professional mountaineer \\
    Military Personnel & Expedition Leader & Professional rock climber \\
    Romantic Couple Campers & Urban Digital Nomad & Pre-teen camper \\
    Wildlife Photographer & Rainforest Explorer & Elderly with arthritis \\
    Field Researchers & High-Altitude Climber & Motion challenged teenager \\
    Solo Budget Traveler & Family Camping Enthusiast & Visually impaired \\
    Mountain Expedition Guide & Emergency Preparedness Advocate & Hearing impaired \\
    Adventure Racer & Festival-goer & Biologist \\
    Scout Leader & Field Researcher & Financially challenged \\
    Eco-Tourism Entrepreneur & Pet-loving Camper & Parent with young children \\
    Extreme Weather Researcher & Urban Activist & Jungle trekker \\
    Long-Distance Cyclist & Van Life Enthusiast & Summer arctic explorer \\
    Humanitarian Worker & Humanitarian Worker & Amputee camper \\
    High School Teacher & Outdoor Educator & Wheelchair accessible camper \\
    Arctic Explorer & Solo Backpacker & Beach camper \\
    Minimalist Enthusiast & Eco-conscious Camper & Back-country portage camper \\
    Wildlife Conservation Volunteer & Outdoor Sports Organizer & Ultramarathon runner \\
    \bottomrule
  \end{tabular}
  \caption{List of user agents generated for each condition.}
  \label{tab:1}
\end{table*}

\subsection{Experiment Setup}

We set up three conditions to test the effectiveness of Elicitron.\ Each condition generated 20 user agents, the same number of people interviewed in~\cite{lin2007}.
\begin{itemize}[noitemsep]
    \item \textbf{Condition 1:} Automatic creation of user agents with the serial method
    \item \textbf{Condition 2:}  Automatic creation of user agents with the serial method and addition of a steering prompt
    \item \textbf{Condition 3:}  Manual creation of ELU agents 
\end{itemize}

The serial method is used because it has been shown in Experiment 1 to create more diverse user agents.\ For Condition 2, we provided the following additional prompt to encourage the creator agent to generate ELU agents.\ The text inside the double quotes are taken verbatim from~\cite{lin2007}.

\begin{customquote}
    You must create non-typical users based on the following description of a typical user: ``The typical user would be a weekend camper, 15-30 years old, with very good health and physical fitness, who camps a few times a year.\ The typical usage environment would be a public park or wilderness area, in a generally wooded or grassy environment with warm, sunny weather.'' 
\end{customquote}

For Condition 3, we manually created ELU agents based on the deviations from the experiences of a typical customer in a typical application and usage environment listed in~\cite{lin2007}.\ The list of all user agents generated for each condition is shown in Table~\ref{tab:1}.

For all three conditions, we prompted the agents to engage in simulated product experience scenarios.\ We then asked LLM agents the following interview questions in sequence. These are the same interview questions used in~\cite{lin2007} with human subjects, but with some modification in wording to encourage the agents to provide specific needs and insights related to the question. 

\begin{itemize}[noitemsep]
    \item \textbf{Free style}: “If you were to purchase an ideal tent, what main characteristics would you look for?”
    \item \textbf{Categorical}: “Focusing specifically on the [category], aspect of tent, can you tell me your needs and any innovative insights to address those needs?”

\textit{Categories:} {size, shape, weight, material, safety, durability, aesthetics, ergonomics, cost, setup, transport} \label{C}
\end{itemize}

\subsection{Latent Needs Labeling}

The responses given by the user agents were analyzed to identify the number of latent needs suggested by each agent.\ Again, we followed the criteria used by~\cite{lin2007} to label whether a particular phrase was a latent need or not:

\begin{customquote}
        If a reported customer need represented a significant change to the product design and did not match the categories [used in interview questions], then it was labeled as a latent need.\ Latent needs were also identified when a reported customer need represented an innovative insight into the product and/or product usage conditions.
\end{customquote}


Two raters performed the labeling task.\ Because determining a latent need is highly subjective and context-dependent, the labeling was performed as follows.\ We began by randomly selecting 10\% of the dataset for independent labeling by both raters.\ Following a calibration discussion to resolve discrepancies, we randomly selected another 20\% of the data to calculate the inter-rater agreement score.\ Finally, the remaining 70\% of the dataset was divided equally between the raters for independent labeling.

Because identifying latent needs from a free text is equivalent to an information retrieval task, an F-score is used to measure the inter-rate agreement as suggested by~\cite{hripcsak2005agreement}.\ The F-score is computed as
\begin{equation}
F_1 = 2 tp / (2 tp + fp + fn)
\end{equation}

If both raters agreed on a particular phrase in the text as a latent need, it was counted as a true positive, $tp$.\ If the first rater identified it as a latent need but not the second rater, it was counted as a false positive, $fp$.\ If the second rater identified it as a latent need but not the first rater, it was counted as a false negative, $fn$.\ We found the F-score of 0.83 ($tp=109$, $fp=21$, $fn=24$), which indicated reliable agreement between the raters.

\subsection{Experiment Results}

\begin{figure*}[ht]
    \centering
    \includegraphics[width=0.9\textwidth]{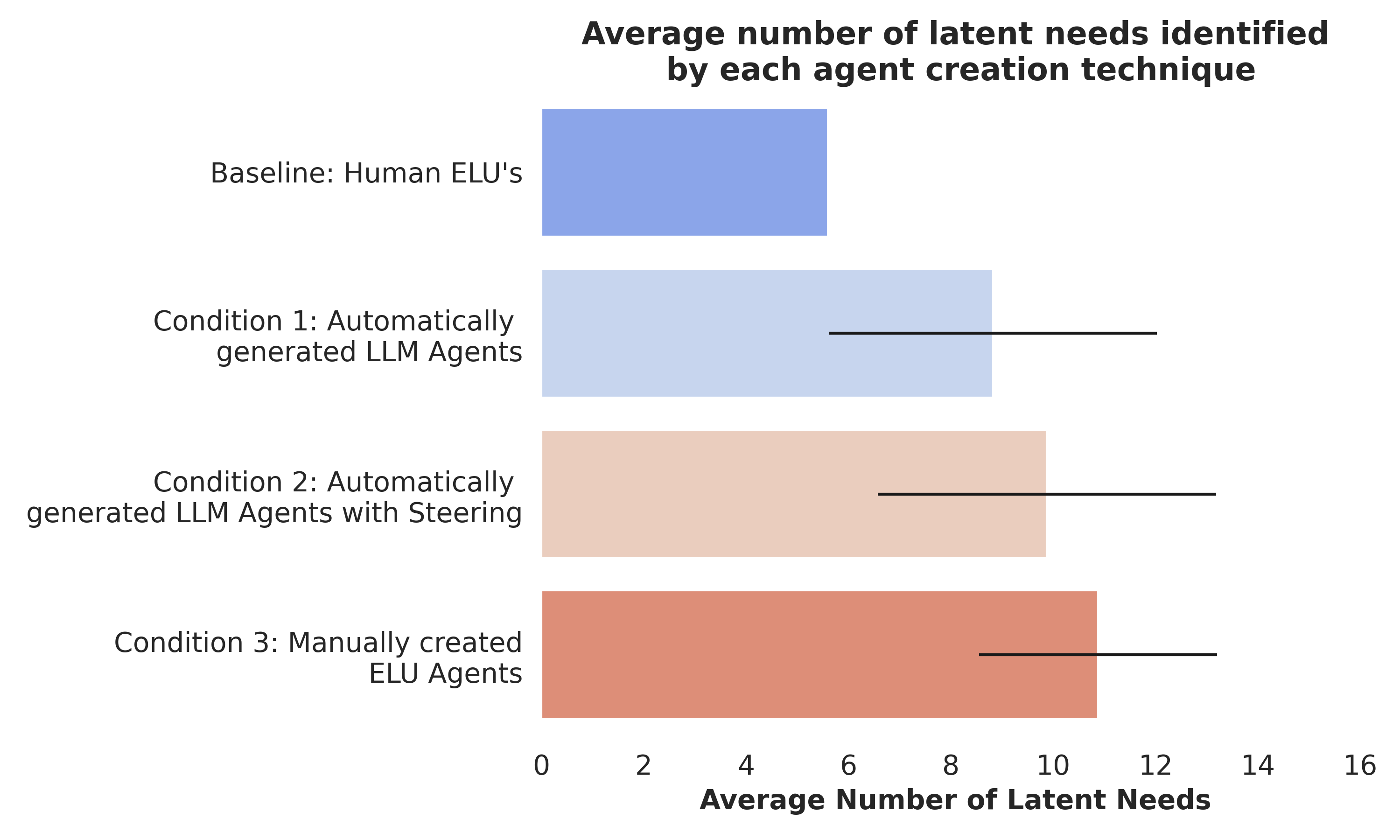}
    \caption{Comparison of the average number of latent needs identified by each user agent across the experimental conditions.\ The error bars indicate standard deviation with n=20 for each condition.}
    \label{fig:enter-label}
\end{figure*}

The results (Figure~\ref{fig:enter-label}) show that the number of latent needs identified was higher for all three Elicitron conditions compared to the baseline, demonstrating the potential of our LLM-based requirements elicitation framework in identifying latent needs.\ Because the prior work~\cite{lin2007} reported the mean latent needs but not any measure of variance, we could not conduct any statistical test to show statistical significance. 

Among the three Elicitron conditions, Condition 3, the manual creation of ELU agents, led to the most number of needs identified (M = 10.875, SD = 2.322), followed by Condition 2, the automatic creation with the steering prompt (M = 9.875, SD = 3.304), and Condition 1, the automatic creation only (M = 8.825, SD = 3.201). A statistically significant difference was found between Condition 3 and Condition 1, t(38) = 2.318, p < 0.05. 

While Condition 1 still produced a fairly diverse set of user agents as shown in Table \ref{tab:1}, most of them do not necessarily represent ELU agents that could lead to the identification of latent needs.\ For example, while user agents such as ``Adventure Racer'', ``Eco-Tourism Entrepreneur'', or ``High School Teacher'' could be considered unique user types, they did not entail extraordinary usage conditions and therefore, led to the identification of relatively few latent needs (6, 3, and 4, respectively).

The difference in the average number of latent needs between Condition 3 and Condition 2 was not statistically significant.\ This indicates that an effective steering prompt can eliminate the need to manually create ELU agents.\ In fact, many of the user agents created in Condition 2 are similar to the ELU agents manually created in Condition 3. Examples include those who are outside of the typical ages (``Adventure-seeking Teen'', ``Retired Nature Enthusiast''), those who are physically challenged (``Person with Physical Disability''), those facing extraordinary conditions (``Expedition Leader'', ``High-Altitude Climber''), etc.


\subsection{Example Outputs from LLM Agents}
\label{sec:example_output}
Example outputs produced by LLM agents in the experiment are shown here.\ 
First, the product experience generated for a user agent titled ``Elderly with arthritis'':

\begin{customquote}
Step 1: 
\begin{itemize}[noitemsep,topsep=0pt]
    \item Action: Attempted to grasp the zipper of the tent’s entrance with limited finger dexterity.
    \item Observation: The zipper was too small and lacked grip, making it difficult to manipulate.
    \item Challenge: Faced significant struggle to open and close the tent, leading to frustration. 
\end{itemize}
Step 2:
\begin{itemize}[noitemsep,topsep=0pt]
    \item Action: Tried to assemble the tent poles and stretch the fabric over the structure. 
    \item Observation: The effort required to connect the tent poles and stretch the fabric exacerbated my arthritis pains.
    \item Challenge: Experienced severe difficulty due to lack of strength and pain in fingers, making assembly a painful and time-consuming process.
\end{itemize}
Step 3:
\begin{itemize}[noitemsep,topsep=0pt]
    \item Action: Attempted to stake the tent to the ground.
    \item Observation: Found the standard stakes and hammering method cumbersome and impossible to manage with my condition.
    \item Challenge: Couldn't effectively secure the tent to the ground, raising safety concerns in windy conditions.
\end{itemize}
Step 4:
\begin{itemize}[noitemsep,topsep=0pt]
    \item Action: Tried to pack away the tent after use.
    \item Observation: Struggled to fold the tent and roll it tightly enough to fit back into its bag.
    \item Challenge: The process was physically demanding and required more dexterity and strength than I possess, leading to reliance on others for assistance.
\end{itemize}

\end{customquote}
Here are example interview responses given by the same agent:

\begin{customquote}
\textbf{Q}: If you were to purchase an ideal tent, what main characteristics would you look for?
    \begin{itemize}[noitemsep,topsep=0pt]
        \item ``The ideal tent would need to have large, easy-grip zippers designed for individuals with limited finger dexterity.\ It would include pre-assembled or quick-assembly poles that don't require significant strength or intricate manipulation, possibly with a simple click-together mechanism.\ Additionally, the tent should feature a staking system that doesn't rely on traditional hammering, perhaps using twist-in stakes that are more arthritis-friendly.''
    \end{itemize}
\smallskip
\textbf{Q}: Focusing specifically on the setup aspect of tent, can you tell me your needs and any innovative insights to address those needs?
\begin{itemize}[noitemsep,topsep=0pt]
\item ``... a self-erecting tent structure where the tent automatically expands and sets itself up upon being removed from its bag, eliminating the need for manually connecting tent poles or stretching fabric.\ This could leverage spring-loaded or memory material technology, where the structural elements are designed to automatically assume the correct form and tension when unleashed.''
\end{itemize}
\end{customquote}
Lastly, here are some examples of interesting latent needs identified from the experiment:

\begin{customquote}
\textbf{User}: Visually impaired
\begin{itemize}[noitemsep,topsep=0pt,leftmargin=*]
    \item  ``This could mean a tent with a base that subtly slopes down towards the door, paired with a distinctive tactile path on the floor that leads directly to the entrance/exit.''
\end{itemize}
\smallskip
\textbf{User}: Wheelchair accessible camper
\begin{itemize}[noitemsep,topsep=0pt,leftmargin=*]
    \item  ``All tent controls, such as zippers, vents, and lighting, should be within easy reach from a seated position.''
\end{itemize}
\smallskip
\textbf{User}: Hunter (needs to set up a tent in dark)
\begin{itemize}[noitemsep,topsep=0pt,leftmargin=*]
    \item  ``... integrates a temporary, battery-powered LED guidance system. This system would activate upon initiating the setup process, illuminating each component in sequence (e.g., poles, connectors, and fabric) and guiding the user through the steps for assembly.''
\end{itemize}
\smallskip
\textbf{User}: High-Altitude Climber
\begin{itemize}[noitemsep,topsep=0pt,leftmargin=*]
    \item  ``For enhanced stability in diverse conditions, the development of an adaptive anchoring system that automatically adjusts tension in response to wind and snow conditions could be revolutionary.''
\end{itemize}
\smallskip
\textbf{User}: Outdoor sports organizer
\begin{itemize}[noitemsep,topsep=0pt,leftmargin=*]
    \item  ``A modular design ... would allow for connecting multiple tent units easily to expand the covered area ... the incorporation of a seamless interlocking system that enables tents to be connected without gaps or weak points.''
\end{itemize}

\end{customquote}

\section{Experiment 3: Automatic Detection of Latent User Needs}
Analyzing interviews to detect latent needs is a challenging and time-consuming task that requires a deep understanding of the product and the customer's requirements.\ It can consume valuable resources and may not always yield consistent results across different analysts.

\subsection{Experiment Setup}
In this experiment, we evaluate the performance of LLMs in automating the analysis of interview texts and detecting latent needs.\ To facilitate this, we create a dataset consisting of 20 latent needs and 20 non-latent needs, based on the evaluations of human experts in Experiment 2.\ This dataset will be used to assess the LLM's ability to accurately identify latent needs.

We conduct the evaluation using three different approaches:

\begin{enumerate} \item \textbf{Zero-shot Detection:} The LLM is tasked with labeling latent needs without any additional context, simply by answering the question ``Is this a latent need?'' with a binary response (True or False).\ There is no extra information or criteria given.\ The LLM relies solely on its existing knowledge to answer ``True'' or ``False''.
\item \textbf{Detection with Latent Needs Criteria:} In this approach, the LLM is provided with the criteria to evaluate latent needs.\ Given these criteria, the LLM is asked to answer the question "Is this a latent need?" with a binary response (True or False).\ The criteria are as follows (adopted from~\cite{lin2007}), which are the same criteria used by the human evaluators in Experiment 2:

\begin{customquote}
``Label the reported customer need as a latent need (latent = True) if either of the following conditions is met:
\begin{enumerate}
\item The need represents a significant change to the product design and does not fall into any of the following categories: size, shape, weight, material, safety, durability, aesthetics, ergonomics, cost, setup, or transport.
\item The need reflects an exceptionally innovative and clearly expressed insight regarding the product and/or how it is used.''
\end{enumerate}
\end{customquote}

\item \textbf{Detection with Latent Needs Criteria and Chain-of-Thought:} In this final approach, the LLM is provided with the same criteria for latent needs as in the previous evaluation. However, the LLM is also instructed to use chain-of-thought analysis and think step-by-step to detect the latent needs.\ Given the criteria and the output of the chain-of-thought, the LLM answers the question ``Is this a latent need?'' with a binary response (True or False).
\end{enumerate}

\subsection{Experiment Results}
\begin{figure*}[ht!]
    \centering
    \begin{subfigure}[b]{0.32\textwidth}
        \includegraphics[width=\textwidth]{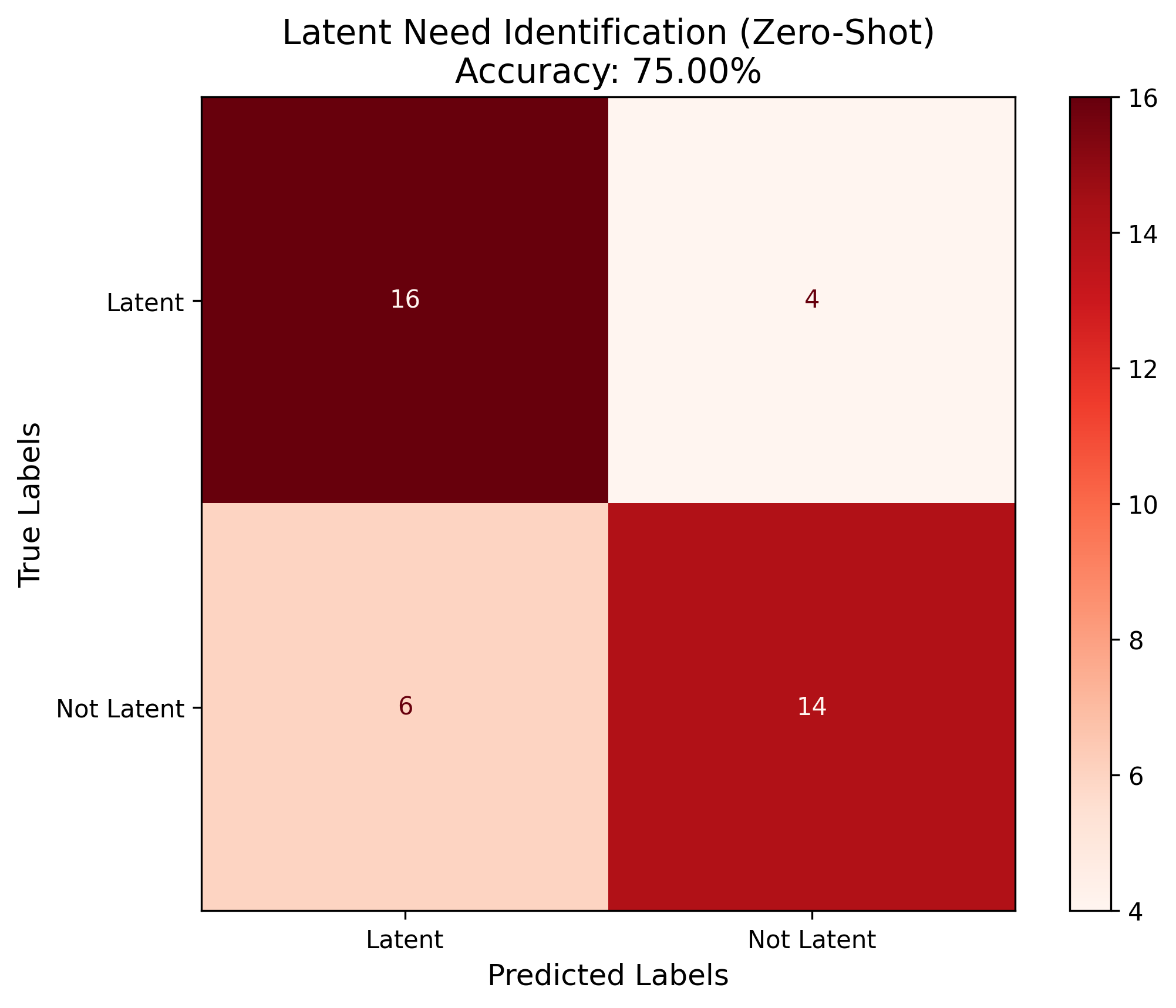}
        \caption{Zero-Shot}
        \label{fig:zero_shot}
    \end{subfigure}
    \hfill 
    \begin{subfigure}[b]{0.32\textwidth}
        \includegraphics[width=\textwidth]{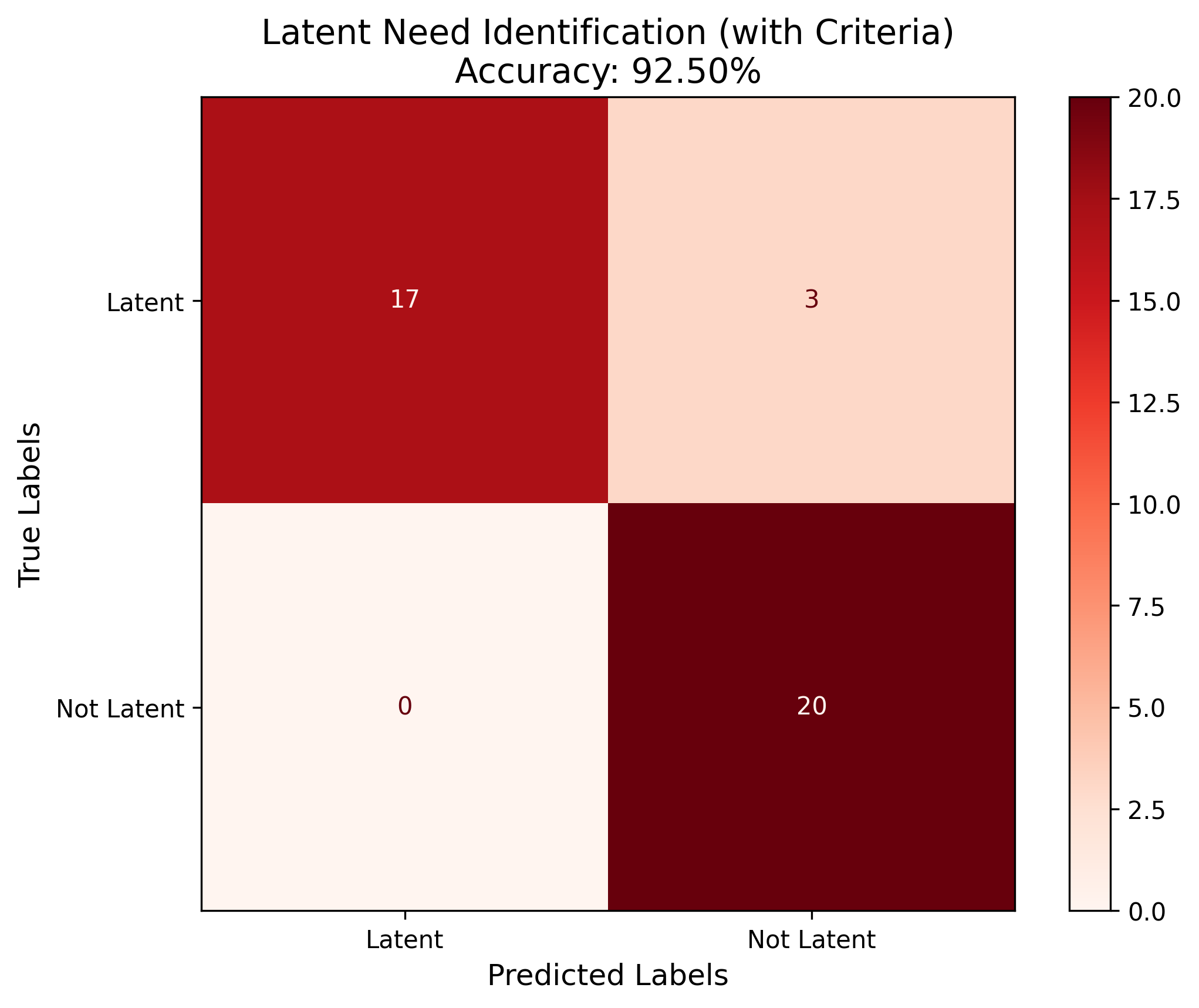}
        \caption{With Criteria}
        \label{fig:with_criteria}
    \end{subfigure}
    \hfill 
    \begin{subfigure}[b]{0.32\textwidth}
        \includegraphics[width=\textwidth]{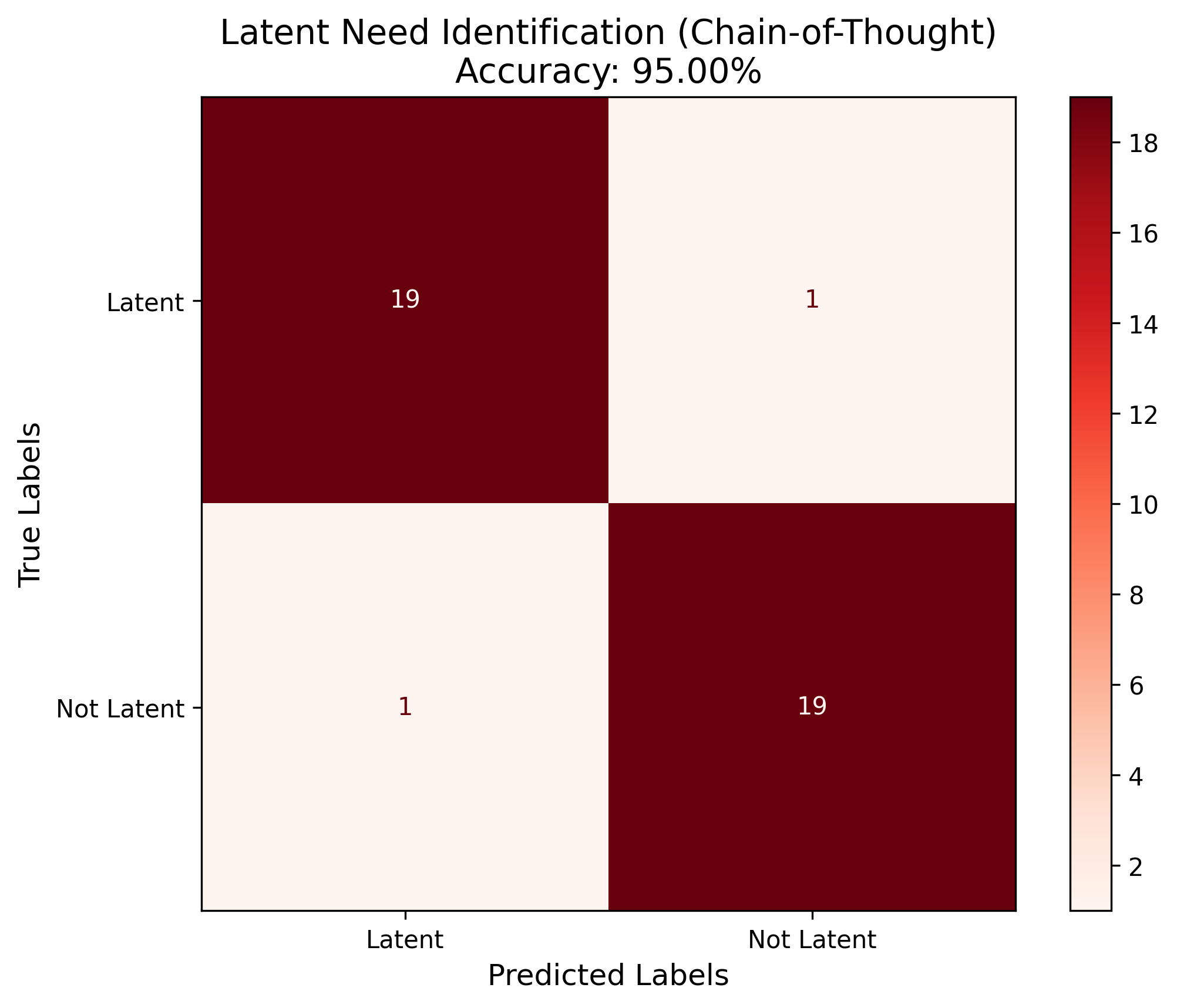}
        \caption{Chain-of-Thought}
        \label{fig:chain_of_thought}
    \end{subfigure}
    \caption{Comparative confusion matrices for latent need identification: (a) zero-Shot classification, (b) classification with latent need criteria, and (c) classification employing a chain-of-thought approach and latent need criteria.}
    \label{fig:confusion_matrices}
\end{figure*}

Figure \ref{fig:confusion_matrices} presents the confusion matrices for the three evaluations, while Table \ref{tab:table_exp3} shows the corresponding performance metrics.\ In the zero-shot detection scenario, the LLM achieves a precision of 0.7273, recall of 0.8000, and an F1-score of 0.7619. These metrics indicate that the LLM can identify latent needs with reasonable accuracy purely based on their internal representations, even without any additional context.\ However, there is room for improvement, as the model struggles to distinguish between latent and non-latent needs in several cases.

When provided with the latent needs criteria, the LLM's performance improves significantly.\ The precision reaches 1.0000, indicating that all the needs identified as latent by the LLM are indeed latent needs. The recall increases to 0.8500, suggesting that the LLM can identify a higher proportion of the actual latent needs in the dataset.

The most impressive results are observed when the LLM is instructed to use chain-of-thought analysis in combination with the latent needs criteria.\ In this case, the precision, recall, and F1-score all reach 0.9500.\ This indicates that the LLM can accurately identify latent needs while minimizing both false positives and false negatives. 

The observed improvements in the LLM's performance across the three scenarios are logical and expected.\ Each successive approach provides the model with increasingly relevant context and guidance, enabling it to better address the task at hand.

First, by providing the latent needs criteria, the LLM is able to focus its attention on the specific aspects that determine whether a need is latent or not.\ This targeted context helps the model zero in on the most pertinent information for making its determination.

Building upon that, the chain-of-thought analysis takes things a step further by guiding the LLM through a structured reasoning process.\ By breaking down the analysis into a series of smaller, interconnected steps, the model is able to systematically consider each criterion and build a logical argument for its ultimate conclusion.\ This approach helps ensure the LLM's output is well-reasoned and grounded in the provided criteria.

The effectiveness of this approach is illustrated in Table~\ref{tab:cot_latent_need}, which presents two examples of the LLM's chain-of-thought reasoning for a latent and a non-latent need.\ In each case, the model's step-by-step analysis clearly demonstrates how it arrived at its determination by carefully considering and applying each of the latent needs criteria in turn.

Beyond just improving performance, the use of chain-of-thought reasoning also enhances the interpretability of the model.\ Chain-of-thought opens a window into the LLM's decision-making process, and makes it possible to understand the apparent logic behind each of the model's conclusions.\ This transparency is valuable in building trust and confidence in the model's outputs, as it allows users to verify that the LLM's decisions are based on sound logic and adhere to the specified criteria.

\begin{table*}[h!]
  \centering
  \begin{tabular}{lSSS}
    \toprule
    \textbf{Matrix} & \textbf{Precision} & \textbf{Recall} & \textbf{F1-score} \\
    \midrule
    Zero-Shot & 0.7273 & 0.8000 & 0.7619 \\
    With Criteria & 1.0000 & 0.8500 & 0.9189 \\
    Chain-of-Thought & 0.9500 & 0.9500 & 0.9500 \\
    \bottomrule
  \end{tabular}
  \caption{Precision, Recall, and F1-score for Confusion Matrices}
  \label{tab:table_exp3}
\end{table*}

\begin{table*}[htbp]
\centering
\begin{tabularx}{\textwidth}{X X}
\toprule
\textbf{Latent Need} & \textbf{Non-Latent Need} \\
\midrule
\textbf{Interview:} For the interior, a design focused on wide angles would maximize the field of view for photography through designated openings, enhancing the potential for wildlife observation without the need to leave the tent.\ & \textbf{Interview:} Regarding the durability aspect of tents, my primary need is for a tent floor that is resilient against tears from sharp objects and general wear. \\
\midrule
\textbf{LLM's chain-of-thought:} The customer need specifies a significant redesign focusing on the tent's interior to maximize the field of view for photography, specifically through designated openings.\ This change goes beyond merely adjusting the size, shape, weight, material, safety, durability, aesthetics, ergonomics, cost, setup, or transport of the tent.\ It addresses an innovative use of the tent interior for wildlife observation through photography that does not necessarily involve leaving the tent, which represents a unique insight into how tents can be utilized.\ This need clearly falls under the first criterion for being considered a latent need, as it is a significant change to the product design that does not fit into the specified exclusion categories.\ Furthermore, it reflects an exceptionally innovative and clearly expressed insight regarding the product and how it is used, making it a valid example of a latent need.\ &
\textbf{LLM's chain-of-thought:} The customer's need is centered on enhancing the durability of the tent floor, specifically making it resistant to tears from sharp objects and general wear.\ This need falls directly into one of the predefined categories (durability) that does not qualify as a latent need according to the stated criteria.\ Therefore, despite being a valid concern and a potential area for product improvement, this need does not represent a significant deviation from existing product design paradigms or introduce an exceptionally innovative insight.\ Thus, it does not meet the conditions for being labeled as a latent need. \\
\bottomrule
\end{tabularx}
\caption{Examples chain-of-thought reasoning by the LLM for latent need identification}
\label{tab:cot_latent_need}
\end{table*}

\section{Cost Analysis}
Cost is an important consideration in latent needs identification studies, and the use of LLMs can be a cost-effective alternative. For example, \textit{GPT-4-Turbo}, which was used in this study, costs 10 USD per 1 million input tokens and 30 USD per 1 million output tokens. This resulted in minimal costs for Experiments 1 and 3, amounting to just a few cents, while Experiment 2, which involved generating approximately 80,000 tokens, incurred a total cost of around 2.4 USD. These costs are significantly lower than those associated with traditional user studies and surveys, which often require substantial investments in participant recruitment, compensation, and data analysis.\ The cost-effectiveness of using LLMs for latent needs identification has important implications for the scalability and accessibility of this research, enabling researchers and organizations with limited budgets to conduct comprehensive studies and obtain valuable insights at a fraction of the cost of conventional methods.

\section{Conclusions}
This paper presents Elicitron, a framework that leverages LLMs to enhance requirements elicitation and uncover diverse user needs, including those that are latent.\ Our context-aware serial generation method proved most effective in creating diverse user agents.\ Elicitron successfully outperformed traditional empathic lead user interviews in generating latent needs. LLMs also demonstrate effectiveness in analyzing interviews and classifying latent needs. Elicitron shows the potential of leveraging LLMs for requirements elicitation and user-centered design, providing an alternative and cost-effective method for designers.

While Elicitron shows promise, the quality of insights depends on the LLM's capabilities, and prioritizing latent needs remains a designer's task.\ Future work includes user studies to validate Elicitron's ability to aid designers and exploring multi-agent interactions to uncover broader unmet needs.\ Moreover, exploring the possibility of incorporating multimodal inputs and outputs in the process of eliciting requirements represents another prospective avenue for research.

\bibliographystyle{asmeconf}  
\bibliography{main}

\begin{thebibliography}{10}
\newcommand{\enquote}[1]{``#1''}
\providecommand{\url}[1]{\texttt{#1}}
\providecommand{\urlprefix}{URL }
\expandafter\ifx\csname urlstyle\endcsname\relax
  \providecommand{\doi}[1]{DOI \discretionary{}{}{}#1}\else
  \providecommand{\doi}{DOI \discretionary{}{}{}\begingroup \urlstyle{rm}\Url}\fi
\providecommand{\eprint}[2][]{\urlprefix\url{#1#2}}

\bibitem{zave1997classification}
Zave, Pamela.
\newblock \enquote{Classification of research efforts in requirements engineering.}
\newblock \textit{ACM Computing Surveys (CSUR)} Vol.~29 No.~4 (1997): pp. 315--321.

\bibitem{berry2007ambiguity}
Berry, Daniel~M.
\newblock \enquote{Ambiguity in natural language requirements documents.}
\newblock \textit{Monterey Workshop}: pp. 1--7. 2007. Springer.

\bibitem{brown2020language}
Brown, Tom, Mann, Benjamin, Ryder, Nick, Subbiah, Melanie, Kaplan, Jared~D, Dhariwal, Prafulla, Neelakantan, Arvind, Shyam, Pranav, Sastry, Girish, Askell, Amanda et~al.
\newblock \enquote{Language models are few-shot learners.}
\newblock \textit{Advances in neural information processing systems} Vol.~33 (2020): pp. 1877--1901.

\bibitem{wei2022chain}
Wei, Jason, Wang, Xuezhi, Schuurmans, Dale, Bosma, Maarten, Xia, Fei, Chi, Ed, Le, Quoc~V, Zhou, Denny et~al.
\newblock \enquote{Chain-of-thought prompting elicits reasoning in large language models.}
\newblock \textit{Advances in Neural Information Processing Systems} Vol.~35 (2022): pp. 24824--24837.

\bibitem{maalej2015toward}
Maalej, Walid, Nayebi, Maleknaz, Johann, Timo and Ruhe, Guenther.
\newblock \enquote{Toward data-driven requirements engineering.}
\newblock \textit{IEEE software} Vol.~33 No.~1 (2015): pp. 48--54.

\bibitem{lin2007}
Lin, Joseph and Seepersad, Carolyn~Conner.
\newblock \enquote{Empathic Lead Users: The Effects of Extraordinary User Experiences on Customer Needs Analysis and Product Redesign.}
\newblock \textit{International {{Design Engineering Technical Conferences}} and {{Computers}} and {{Information}} in {{Engineering Conference}}}, Vol. 48043: pp. 289--296. 2007.

\bibitem{vaswani2017attention}
Vaswani, Ashish, Shazeer, Noam, Parmar, Niki, Uszkoreit, Jakob, Jones, Llion, Gomez, Aidan~N, Kaiser, {\L}ukasz and Polosukhin, Illia.
\newblock \enquote{Attention is all you need.}
\newblock \textit{Advances in neural information processing systems} Vol.~30 (2017).

\bibitem{lingard2023writing}
Lingard, Lorelei.
\newblock \enquote{Writing with ChatGPT: An illustration of its capacity, limitations \& implications for academic writers.}
\newblock \textit{Perspectives on medical education} Vol.~12 No.~1 (2023): p. 261.

\bibitem{htet2024chatgpt}
Htet, Arkar, Liana, Sui~Reng, Aung, Theingi and Bhaumik, Amiya.
\newblock \enquote{ChatGPT in Content Creation: Techniques, Applications, and Ethical Implications.}
\newblock \textit{Advanced Applications of Generative AI and Natural Language Processing Models}.
\newblock IGI Global (2024): pp. 43--68.

\bibitem{subagja2023improving}
Subagja, Agus~Dedi, Ausat, Abu Muna~Almaududi, Sari, Ade~Risna, Wanof, M~Indre and Suherlan, Suherlan.
\newblock \enquote{Improving Customer Service Quality in MSMEs through the Use of ChatGPT.}
\newblock \textit{Jurnal Minfo Polgan} Vol.~12 No.~2 (2023): pp. 380--386.

\bibitem{li2022competition}
Li, Yujia, Choi, David, Chung, Junyoung, Kushman, Nate, Schrittwieser, Julian, Leblond, R{\'e}mi, Eccles, Tom, Keeling, James, Gimeno, Felix, Dal~Lago, Agustin et~al.
\newblock \enquote{Competition-level code generation with alphacode.}
\newblock \textit{Science} Vol. 378 No. 6624 (2022): pp. 1092--1097.

\bibitem{shanahan2023role}
Shanahan, Murray, McDonell, Kyle and Reynolds, Laria.
\newblock \enquote{Role play with large language models.}
\newblock \textit{Nature} Vol. 623 No. 7987 (2023): pp. 493--498.

\bibitem{csepregi2021effect}
Csepregi, Lajos~Matyas.
\newblock \enquote{The Effect of Context-aware LLM-based NPC Conversations on Player Engagement in Role-playing Video Games.}
\newblock \textit{Unpublished manuscript}  (2021).

\bibitem{zhu2023calypso}
Zhu, Andrew, Martin, Lara, Head, Andrew and Callison-Burch, Chris.
\newblock \enquote{CALYPSO: LLMs as Dungeon Master's Assistants.}
\newblock \textit{Proceedings of the AAAI Conference on Artificial Intelligence and Interactive Digital Entertainment}, Vol.~19. ~1: pp. 380--390. 2023.

\bibitem{gray2015}
Gray, Colin, Yilmaz, Seda, McKilligan, Seda, Daly, Shanna, Seifert, Colleen and Gonzalez, Richard.
\newblock \enquote{Idea Generation through Empathy: {{Reimagining}} the `Cognitive Walkthrough'.}  (2015).

\bibitem{schmitt2016}
Schmitt, Elizabeth and Morkos, Beshoy.
\newblock \enquote{Teaching {{Students Designer Empathy}} in {{Senior Capstone Design}}.}
\newblock \textit{Capstone {{Design Conference}}, {{June}}}: pp. 6--8. 2016.

\bibitem{walther2012}
Walther, Joachim, Miller, Shari~E. and Kellam, Nadia~N.
\newblock \enquote{Exploring the Role of Empathy in Engineering Communication through a Transdisciplinary Dialogue.}
\newblock \textit{2012 {{ASEE Annual Conference}} \& {{Exposition}}}: pp. 25--622. 2012.

\bibitem{hannukainen2006}
Hannukainen, Pia and {Ho{\textbackslash}" ltta{\textbackslash}"-Otto}, Katja.
\newblock \enquote{Identifying Customer Needs: {{Disabled}} Persons as Lead Users.}
\newblock \textit{International Design Engineering Technical Conferences and Computers and Information in Engineering Conference}, Vol. 42584: pp. 243--251. 2006.

\bibitem{Leonard1997}
Leonard, Dorothy and Rayport, Jeffrey~F.
\newblock \enquote{Spark Innovation through Empathic Design.}
\newblock \textit{Harvard business review} Vol.~75 (1997): pp. 102--115.

\bibitem{Otto2003}
Otto, Kevin~N.
\newblock \textit{Product Design: Techniques in Reverse Engineering and New Product Development}.
\newblock {Tsinghua University Press Co., Ltd.} (2003).

\bibitem{Ulrich2016}
Ulrich, Karl~T. and Eppinger, Steven~D.
\newblock \textit{Product Design and Development}.
\newblock {McGraw-hill} (2016).

\bibitem{Suh1990}
Suh, Nam~P.
\newblock \enquote{The Principles of Design: {{Oxford}} University Press.}
\newblock \textit{New York}  (1990).

\bibitem{vonhippel1986}
Von~Hippel, Eric.
\newblock \enquote{Lead {{Users}}: {{A Source}} of {{Novel Product Concepts}}.}
\newblock \textit{Management Science} Vol.~32 No.~7 (1986): pp. 791--805.
\newblock \doi{10.1287/mnsc.32.7.791}.

\bibitem{urban1988}
Urban, Glen~L. and Von~Hippel, Eric.
\newblock \enquote{Lead {{User Analyses}} for the {{Development}} of {{New Industrial Products}}.}
\newblock \textit{Management Science} Vol.~34 No.~5 (1988): pp. 569--582.
\newblock \doi{10.1287/mnsc.34.5.569}.

\bibitem{issa2023}
Issa, Nurhayati~Md, Sasaki, Hayata, Okamura, Nami, Yahya, Wira~Jazair, Rahman, Mohd Azizi~Abdul, Ariff, Mohd Hatta~Mohammed and Koga, Tsuyoshi.
\newblock \enquote{Proposition and {{Verification}} of a {{Design Method}} to {{Discover Latent Needs Based}} on {{Empathy}}, {{Experiences}}, and {{Working Prototype}} by {{Designing Autonomous Childcare Vehicle}}.}
\newblock \textit{Journal of Advanced Vehicle System} Vol.~14 No.~1 (2023): pp. 19--34.

\bibitem{zhu2023}
Zhu, Qihao and Luo, Jianxi.
\newblock \enquote{Toward {{Artificial Empathy}} for {{Human-Centered Design}}: {{A Framework}}.} (2023).
\newblock \doi{10.48550/arXiv.2303.10583}.
\newblock Accessed 2023-12-18, \eprint{2303.10583}.

\bibitem{strobel2013}
Strobel, Johannes, Hess, Justin, Pan, Rui and Wachter~Morris, Carrie~A.
\newblock \enquote{Empathy and Care within Engineering: Qualitative Perspectives from Engineering Faculty and Practicing Engineers.}
\newblock \textit{Engineering Studies} Vol.~5 No.~2 (2013): pp. 137--159.
\newblock \doi{10.1080/19378629.2013.814136}.

\bibitem{raviselvam2017}
Raviselvam, Sujithra, Sanaei, Roozbeh, Blessing, Lucienne, {H{\"o}ltt{\"a}-Otto}, Katja and Wood, Kristin~L.
\newblock \enquote{Demographic Factors and Their Influence on Designer Creativity and Empathy Evoked through User Extreme Conditions.}
\newblock \textit{International {{Design Engineering Technical Conferences}} and {{Computers}} and {{Information}} in {{Engineering Conference}}}, Vol. 58219: p. V007T06A011. 2017. {American Society of Mechanical Engineers}.

\bibitem{surma-aho2018}
{Surma-Aho}, Antti, Bj{\"o}rklund, Tua and {H{\"o}ltt{\"a}-Otto}, Katja.
\newblock \enquote{An Analysis of Designer Empathy in the Early Phases of Design Projects.}
\newblock \textit{DS 91: Proceedings of NordDesign 2018, Link{\"o}ping, Sweden, 14th-17th August 2018}  (2018).

\bibitem{tang2018}
Tang, Xiaofeng.
\newblock \enquote{From'{{Empathic Design}}'to'{{Empathic Engineering}}': {{Toward}} a {{Genealogy}} of {{Empathy}} in {{Engineering Education}}.}
\newblock \textit{2018 {{ASEE Annual Conference}} \& {{Exposition}}}. 2018.

\bibitem{boden2009computer}
Boden, Margaret~A.
\newblock \enquote{Computer models of creativity.}
\newblock \textit{AI Magazine} Vol.~30 No.~3 (2009): pp. 23--23.

\bibitem{amabile1988model}
Amabile, Teresa~M et~al.
\newblock \enquote{A model of creativity and innovation in organizations.}
\newblock \textit{Research in organizational behavior} Vol.~10 No.~1 (1988): pp. 123--167.

\bibitem{amabile1982social}
Amabile, Teresa~M.
\newblock \enquote{Social psychology of creativity: A consensual assessment technique.}
\newblock \textit{Journal of personality and social psychology} Vol.~43 No.~5 (1982): p. 997.

\bibitem{miller2021should}
Miller, Scarlett~R, Hunter, Samuel~T, Starkey, Elizabeth, Ramachandran, Sharath, Ahmed, Faez and Fuge, Mark.
\newblock \enquote{How should we measure creativity in engineering design? A comparison between social science and engineering approaches.}
\newblock \textit{Journal of Mechanical Design} Vol. 143 No.~3 (2021).

\bibitem{amabile2018creativity}
Amabile, Teresa~M.
\newblock \textit{Creativity in context: Update to the social psychology of creativity}.
\newblock Routledge (2018).

\bibitem{regenwetter2023statistical}
Regenwetter, Lyle, Srivastava, Akash, Gutfreund, Dan and Ahmed, Faez.
\newblock \enquote{Beyond Statistical Similarity: Rethinking Metrics for Deep Generative Models in Engineering Design.} (2023).
\newblock \eprint{2302.02913}.

\bibitem{ma2023conceptual}
Ma, Kevin, Grandi, Daniele, McComb, Christopher and Goucher-Lambert, Kosa.
\newblock \enquote{Conceptual Design Generation Using Large Language Models.} (2023).
\newblock \eprint{2306.01779}.

\bibitem{jiralerspong2024feature}
Jiralerspong, Marco, Bose, Joey, Gemp, Ian, Qin, Chongli, Bachrach, Yoram and Gidel, Gauthier.
\newblock \enquote{Feature Likelihood Score: Evaluating the Generalization of Generative Models Using Samples.}
\newblock \textit{Advances in Neural Information Processing Systems} Vol.~36 (2024).

\bibitem{sarica2023innovation}
Sarica, Serhad and Luo, Jianxi.
\newblock \enquote{Innovation Slowdown: Decelerating Concept Creation and Declining Originality in New Technological Concepts.}
\newblock \textit{arXiv preprint arXiv:2303.13300}  (2023).

\bibitem{picard2023dated}
Picard, Cyril, Schiffmann, J{\"u}rg and Ahmed, Faez.
\newblock \enquote{DATED: Guidelines for Creating Synthetic Datasets for Engineering Design Applications.}
\newblock \textit{arXiv preprint arXiv:2305.09018}  (2023).

\bibitem{regenwetter2023counterfactuals}
Regenwetter, Lyle, Abu~Obaideh, Yazan and Ahmed, Faez.
\newblock \enquote{Counterfactuals for design: A model-agnostic method for design recommendations.}
\newblock \textit{International Design Engineering Technical Conferences and Computers and Information in Engineering Conference}, Vol. 87301: p. V03AT03A008. 2023. American Society of Mechanical Engineers.

\bibitem{bagazinski2023shipgen}
Bagazinski, Noah~J and Ahmed, Faez.
\newblock \enquote{ShipGen: A Diffusion Model for Parametric Ship Hull Generation with Multiple Objectives and Constraints.}
\newblock \textit{Journal of Marine Science and Engineering} Vol.~11 No.~12 (2023): p. 2215.

\bibitem{fan2023noise}
Fan, Jiajie, Vuaille, Laure, B{\"a}ck, Thomas and Wang, Hao.
\newblock \enquote{On the Noise Scheduling for Generating Plausible Designs with Diffusion Models.}
\newblock \textit{arXiv preprint arXiv:2311.11207}  (2023).

\bibitem{rousseeuw1987silhouettes}
Rousseeuw, Peter~J.
\newblock \enquote{Silhouettes: a graphical aid to the interpretation and validation of cluster analysis.}
\newblock \textit{Journal of computational and applied mathematics} Vol.~20 (1987): pp. 53--65.

\bibitem{podani2009convex}
Podani, J.
\newblock \enquote{Convex hulls, habitat filtering, and functional diversity: mathematical elegance versus ecological interpretability.}
\newblock \textit{Community Ecology} Vol.~10 No.~2 (2009): pp. 244--250.

\bibitem{mueller2015combining}
Mueller, Caitlin~T and Ochsendorf, John~A.
\newblock \enquote{Combining structural performance and designer preferences in evolutionary design space exploration.}
\newblock \textit{Automation in Construction} Vol.~52 (2015): pp. 70--82.

\bibitem{brown2019quantifying}
Brown, Nathan~C and Mueller, Caitlin~T.
\newblock \enquote{Quantifying diversity in parametric design: a comparison of possible metrics.}
\newblock \textit{AI EDAM} Vol.~33 No.~1 (2019): pp. 40--53.

\bibitem{zanitti2022exploiting}
Zanitti, Michele, S{\o}rensen, Jannick, Terolli, Erisa and Kosta, Sokol.
\newblock \enquote{Exploiting Consumption Diversity and neighbour Similarity Trade-offs in Recommender Systems: a User-Centric Offline Evaluation of Diversity Objectives.}  (2022).

\bibitem{chaudhuri2019advanced}
Chaudhuri, Arpita, Sarma, Monalisa and Samanta, Debasis.
\newblock \enquote{Advanced feature identification towards research article recommendation: a machine learning based approach.}
\newblock \textit{TENCON 2019-2019 IEEE Region 10 Conference (TENCON)}: pp. 7--12. 2019. IEEE.

\bibitem{openai2024gpt4}
OpenAI, Achiam, Josh, Adler, Steven, Agarwal, Sandhini, Ahmad, Lama, Akkaya, Ilge, Aleman, Florencia~Leoni, Almeida, Diogo, Altenschmidt, Janko, Altman, Sam, Anadkat, Shyamal, Avila, Red, Babuschkin, Igor, Balaji, Suchir, Balcom, Valerie, Baltescu, Paul, Bao, Haiming, Bavarian, Mohammad, Belgum, Jeff, Bello, Irwan, Berdine, Jake, Bernadett-Shapiro, Gabriel, Berner, Christopher, Bogdonoff, Lenny, Boiko, Oleg, Boyd, Madelaine, Brakman, Anna-Luisa, Brockman, Greg, Brooks, Tim, Brundage, Miles, Button, Kevin, Cai, Trevor, Campbell, Rosie, Cann, Andrew, Carey, Brittany, Carlson, Chelsea, Carmichael, Rory, Chan, Brooke, Chang, Che, Chantzis, Fotis, Chen, Derek, Chen, Sully, Chen, Ruby, Chen, Jason, Chen, Mark, Chess, Ben, Cho, Chester, Chu, Casey, Chung, Hyung~Won, Cummings, Dave, Currier, Jeremiah, Dai, Yunxing, Decareaux, Cory, Degry, Thomas, Deutsch, Noah, Deville, Damien, Dhar, Arka, Dohan, David, Dowling, Steve, Dunning, Sheila, Ecoffet, Adrien, Eleti, Atty, Eloundou, Tyna, Farhi, David, Fedus, Liam, Felix, Niko, Fishman, Simón~Posada, Forte, Juston, Fulford, Isabella, Gao, Leo, Georges, Elie, Gibson, Christian, Goel, Vik, Gogineni, Tarun, Goh, Gabriel, Gontijo-Lopes, Rapha, Gordon, Jonathan, Grafstein, Morgan, Gray, Scott, Greene, Ryan, Gross, Joshua, Gu, Shixiang~Shane, Guo, Yufei, Hallacy, Chris, Han, Jesse, Harris, Jeff, He, Yuchen, Heaton, Mike, Heidecke, Johannes, Hesse, Chris, Hickey, Alan, Hickey, Wade, Hoeschele, Peter, Houghton, Brandon, Hsu, Kenny, Hu, Shengli, Hu, Xin, Huizinga, Joost, Jain, Shantanu, Jain, Shawn, Jang, Joanne, Jiang, Angela, Jiang, Roger, Jin, Haozhun, Jin, Denny, Jomoto, Shino, Jonn, Billie, Jun, Heewoo, Kaftan, Tomer, Łukasz Kaiser, Kamali, Ali, Kanitscheider, Ingmar, Keskar, Nitish~Shirish, Khan, Tabarak, Kilpatrick, Logan, Kim, Jong~Wook, Kim, Christina, Kim, Yongjik, Kirchner, Jan~Hendrik, Kiros, Jamie, Knight, Matt, Kokotajlo, Daniel, Łukasz Kondraciuk, Kondrich, Andrew, Konstantinidis, Aris, Kosic, Kyle, Krueger, Gretchen, Kuo, Vishal, Lampe, Michael, Lan, Ikai, Lee, Teddy, Leike, Jan, Leung, Jade, Levy, Daniel, Li, Chak~Ming, Lim, Rachel, Lin, Molly, Lin, Stephanie, Litwin, Mateusz, Lopez, Theresa, Lowe, Ryan, Lue, Patricia, Makanju, Anna, Malfacini, Kim, Manning, Sam, Markov, Todor, Markovski, Yaniv, Martin, Bianca, Mayer, Katie, Mayne, Andrew, McGrew, Bob, McKinney, Scott~Mayer, McLeavey, Christine, McMillan, Paul, McNeil, Jake, Medina, David, Mehta, Aalok, Menick, Jacob, Metz, Luke, Mishchenko, Andrey, Mishkin, Pamela, Monaco, Vinnie, Morikawa, Evan, Mossing, Daniel, Mu, Tong, Murati, Mira, Murk, Oleg, Mély, David, Nair, Ashvin, Nakano, Reiichiro, Nayak, Rajeev, Neelakantan, Arvind, Ngo, Richard, Noh, Hyeonwoo, Ouyang, Long, O'Keefe, Cullen, Pachocki, Jakub, Paino, Alex, Palermo, Joe, Pantuliano, Ashley, Parascandolo, Giambattista, Parish, Joel, Parparita, Emy, Passos, Alex, Pavlov, Mikhail, Peng, Andrew, Perelman, Adam, de~Avila Belbute~Peres, Filipe, Petrov, Michael, de~Oliveira~Pinto, Henrique~Ponde, Michael, Pokorny, Pokrass, Michelle, Pong, Vitchyr~H., Powell, Tolly, Power, Alethea, Power, Boris, Proehl, Elizabeth, Puri, Raul, Radford, Alec, Rae, Jack, Ramesh, Aditya, Raymond, Cameron, Real, Francis, Rimbach, Kendra, Ross, Carl, Rotsted, Bob, Roussez, Henri, Ryder, Nick, Saltarelli, Mario, Sanders, Ted, Santurkar, Shibani, Sastry, Girish, Schmidt, Heather, Schnurr, David, Schulman, John, Selsam, Daniel, Sheppard, Kyla, Sherbakov, Toki, Shieh, Jessica, Shoker, Sarah, Shyam, Pranav, Sidor, Szymon, Sigler, Eric, Simens, Maddie, Sitkin, Jordan, Slama, Katarina, Sohl, Ian, Sokolowsky, Benjamin, Song, Yang, Staudacher, Natalie, Such, Felipe~Petroski, Summers, Natalie, Sutskever, Ilya, Tang, Jie, Tezak, Nikolas, Thompson, Madeleine~B., Tillet, Phil, Tootoonchian, Amin, Tseng, Elizabeth, Tuggle, Preston, Turley, Nick, Tworek, Jerry, Uribe, Juan Felipe~Cerón, Vallone, Andrea, Vijayvergiya, Arun, Voss, Chelsea, Wainwright, Carroll, Wang, Justin~Jay, Wang, Alvin, Wang, Ben, Ward, Jonathan, Wei, Jason, Weinmann, CJ, Welihinda, Akila, Welinder, Peter, Weng, Jiayi, Weng, Lilian, Wiethoff, Matt, Willner, Dave, Winter, Clemens, Wolrich, Samuel, Wong, Hannah, Workman, Lauren, Wu, Sherwin, Wu, Jeff, Wu, Michael, Xiao, Kai, Xu, Tao, Yoo, Sarah, Yu, Kevin, Yuan, Qiming, Zaremba, Wojciech, Zellers, Rowan, Zhang, Chong, Zhang, Marvin, Zhao, Shengjia, Zheng, Tianhao, Zhuang, Juntang, Zhuk, William and Zoph, Barret.
\newblock \enquote{GPT-4 Technical Report.} (2024).
\newblock \eprint{2303.08774}.

\bibitem{van2008visualizing}
Van~der Maaten, Laurens and Hinton, Geoffrey.
\newblock \enquote{Visualizing data using t-SNE.}
\newblock \textit{Journal of machine learning research} Vol.~9 No.~11 (2008).

\bibitem{hripcsak2005agreement}
Hripcsak, George and Rothschild, Adam~S.
\newblock \enquote{Agreement, the f-measure, and reliability in information retrieval.}
\newblock \textit{Journal of the American medical informatics association} Vol.~12 No.~3 (2005): pp. 296--298.

\end{thebibliography}

\end{document}